\documentclass[default,iicol]{sn-jnl}

\usepackage{graphicx}\usepackage{multirow}\usepackage{amsmath,amssymb,amsfonts}\usepackage{amsthm}\usepackage{mathrsfs}\usepackage[title]{appendix}\usepackage{xcolor}\usepackage{textcomp}\usepackage{manyfoot}\usepackage{booktabs}\usepackage{algpseudocode}\usepackage{listings}

\usepackage[caption=false]{subfig}
\usepackage{pifont}

\usepackage{paralist}
\usepackage{url}

\usepackage[linesnumbered,ruled]{algorithm2e}

\usepackage{pgfplotstable}
\usepackage{xspace}
\usepackage{multirow}
\usepackage{datatool}
\usepackage[gen]{eurosym}
\usepackage[binary-units=true,detect-weight=true, detect-family=true]{siunitx}
\DeclareSIUnit{\pp}{\textup{pp}}

\usepackage[inline]{enumitem}

\newcommand{\approach}{FITI\xspace}

\usepackage{tikz}
\usetikzlibrary{shapes.geometric,arrows}
\usetikzlibrary{positioning,fit,calc}
\usetikzlibrary{arrows.meta}
\tikzstyle{arrow}=[draw]
\usetikzlibrary{arrows}

\usepackage{newtx}
\usepackage{hyperref}
\usepackage[hyphenbreaks]{breakurl}

\theoremstyle{thmstyleone}

\theoremstyle{thmstyletwo}

\theoremstyle{thmstylethree}

\begin{document}

\title[Article Title]{Tracing Content Requirements in Financial Documents using Multi-granularity Text Analysis}

\author*[1]{\fnm{Xiaochen} \sur{Li}}\email{xiaochen.li@dlut.edu.cn}
\equalcont{Part of this work was done while the author was affiliated with
	the University of Luxembourg, Luxembourg.}

\author[2]{\fnm{Domenico} \sur{Bianculli}}\email{domenico.bianculli@uni.lu}

\author[3,4]{\fnm{Lionel} \sur{Briand}}\email{lbriand@uottawa.ca}
\equalcont{Part of this work was done while the author was affiliated with
	the University of Luxembourg, Luxembourg.}

\affil*[1]{\orgname{Dalian University of Technology}, \orgaddress{\city{Dalian}, \postcode{116621}, \state{Liaoning}, \country{China}}}

\affil[2]{\orgname{University of Luxembourg}, \orgaddress{\city{Kirchberg}, \postcode{L-1855}, \country{Luxembourg}}}

\affil[3]{\orgname{Research Ireland Lero Centre for Software Research and University of Limerick}, \orgaddress{\city{Limerick}, \postcode{V94 T9PX}, \country{Ireland}}}

\affil[4]{\orgname{University of Ottawa}, \orgaddress{\city{Ottawa}, \postcode{K1H 8M5}, \country{Canada}}}

\abstract{The completeness (in terms of content) of financial
	documents is a fundamental requirement for investment funds.
	To ensure completeness, financial regulators have to spend significant
	time carefully checking every financial document based on relevant
	content requirements, which prescribe the information types to be
	included in financial documents (e.g., the fund name, the description
	of shares' issue conditions and procedures).
	Although several techniques have been proposed to automatically detect
	certain types of information in documents across application
	domains, they provide limited support to help
	regulators automatically identify the text chunks related to financial
	information types, due to the complexity of financial documents and
	the diversity of the sentences typically characterizing  an information type.

	In this paper, we propose \approach to trace
	content requirements in financial documents with multi-granularity text analysis.
	Given a new financial document, \approach first selects a set of candidate sentences for efficient information type identification.
	Then, to rank candidate sentences, \approach uses a combination of rule-based and data-centric
	approaches, by leveraging  information retrieval (IR) and machine
	learning (ML) techniques that analyze the words, sentences, and
	contexts related to an information type.
	Finally, using a list of domain-specific indicator phrases related to
	each information type, a heuristic-based selector, which considers both the
	sentence ranking and domain-specific phrases, determines a list of sentences corresponding to each information type.

	We evaluated \approach by assessing its effectiveness
	in tracing financial content requirements in 100 real-world financial documents.
	Experimental results show that \approach is able to provide accurate identification
	with average precision, recall, and F$_1$-score values of {\FITIPAvg}, {\FITIRAvg}, and {\FITIFAvg}, respectively.
	The overall accuracy of \approach significantly outperforms the best baseline (based on a transformer language model) by 0.266 in terms of F$_1$-score. Furthermore,
	\approach can help regulators detect about 80\% of missing information types in financial documents.}


\begin{filecontents*}{data/rq-accuracy.csv}
ID, P-KW,  P-BERT, P-BERTKW, P-FITI, R-KW,  R-BERT, R-BERTKW, R-FITI, F-KW,  F-BERT, F-BERTKW, F-FITI
T1, 0.100, 0.407,  0.743,    0.855,  0.682, 0.543,  0.558,    0.722,  0.176, 0.466,  0.512,    0.783
T2, 0.553, 0.301,  0.797,    0.788,  0.463, 0.658,  0.353,    0.536,  0.504, 0.413,  0.490,    0.638
T3, 0.597, 0.258,  0.672,    0.784,  0.326, 0.682,  0.209,    0.662,  0.422, 0.375,  0.318,    0.718
T4, 0.772, 0.485,  0.864,    0.882,  0.615, 0.941,  0.403,    0.893,  0.685, 0.640,  0.549,    0.887
T5, 0.669, 0.275,  0.696,    0.812,  0.293, 0.511,  0.190,    0.418,  0.407, 0.358,  0.298,    0.552
Avg,0.539, 0.345,  0.700,    0.824,  0.476, 0.667,  0.343,    0.646,  0.439, 0.450,  0.433,    0.716
\end{filecontents*}
\DTLloaddb{accuracy}{data/rq-accuracy.csv}
\DTLgetvalue{\FITIPLow}{accuracy}{3}{\dtlcolumnindex{accuracy}{P-FITI}}
\DTLgetvalue{\FITIPHigh}{accuracy}{4}{\dtlcolumnindex{accuracy}{P-FITI}}
\DTLgetvalue{\FITIRLow}{accuracy}{5}{\dtlcolumnindex{accuracy}{R-FITI}}
\DTLgetvalue{\FITIRHigh}{accuracy}{4}{\dtlcolumnindex{accuracy}{R-FITI}}

\DTLgetvalue{\FITIPAvg}{accuracy}{6}{\dtlcolumnindex{accuracy}{P-FITI}}
\DTLgetvalue{\FITIRAvg}{accuracy}{6}{\dtlcolumnindex{accuracy}{R-FITI}}
\DTLgetvalue{\FITIFAvg}{accuracy}{6}{\dtlcolumnindex{accuracy}{F-FITI}}

\DTLgetvalue{\KWPAvg}{accuracy}{6}{\dtlcolumnindex{accuracy}{P-KW}}
\DTLgetvalue{\KWRAvg}{accuracy}{6}{\dtlcolumnindex{accuracy}{R-KW}}
\DTLgetvalue{\KWFAvg}{accuracy}{6}{\dtlcolumnindex{accuracy}{F-KW}}

\DTLgetvalue{\BERTPAvg}{accuracy}{6}{\dtlcolumnindex{accuracy}{P-BERT}}
\DTLgetvalue{\BERTRAvg}{accuracy}{6}{\dtlcolumnindex{accuracy}{R-BERT}}

\DTLgetvalue{\BERTKWPAvg}{accuracy}{6}{\dtlcolumnindex{accuracy}{P-BERTKW}}
\DTLgetvalue{\BERTKWRAvg}{accuracy}{6}{\dtlcolumnindex{accuracy}{R-BERTKW}}
\begin{filecontents*}{data/rq-selection.csv}
ID, P-SC, P-TN, P-AVG, P-FITI, R-SC, R-TN, R-AVG, R-FITI, F-SC, F-TN, F-AVG, F-FITI
T1, 0.713, 0.712, 0.713, \textbf{0.855}, 0.740, \textbf{0.749}, 0.744, 0.722, 0.726, 0.730, 0.728, \textbf{0.783}
T2, 0.407, 0.483, 0.401, \textbf{0.788}, 0.440, \textbf{0.635}, 0.440, 0.536, 0.423, 0.549, 0.420, \textbf{0.638}
T3, 0.706, 0.744, 0.636, \textbf{0.784}, 0.642, 0.606, \textbf{0.669}, 0.662, 0.673, 0.668, 0.652, \textbf{0.718}
T4, \textbf{0.924}, 0.860, 0.834, 0.882, 0.814, 0.855, 0.851, \textbf{0.893}, 0.866, 0.857, 0.842, \textbf{0.887}
T5, 0.612, 0.704, 0.468, \textbf{0.812}, 0.351, 0.334, \textbf{0.429}, 0.418, 0.446, 0.453, 0.448, \textbf{0.552}
Avg,0.672, 0.701, 0.610, \textbf{0.824}, 0.597, 0.636, 0.627, \textbf{0.646}, 0.627, 0.651, 0.618, \textbf{0.716}
\end{filecontents*}
\DTLloaddb{selection}{data/rq-selection.csv}
\DTLgetvalue{\TopNPAvg}{selection}{6}{\dtlcolumnindex{selection}{P-TN}}
\DTLgetvalue{\AvgNPAvg}{selection}{6}{\dtlcolumnindex{selection}{P-AVG}}

\begin{filecontents*}{data/rq-inspection.csv}
ID, P,    R,    F
T1, 0.70, 0.70, 0.70
T2, 0.75, 0.90, 0.82
T3, 0.90, 0.90, 0.90
T4, 1.00, 0.90, 0.95
T5, 0.83, 0.50, 0.63
Avg., 0.84, 0.78, 0.80
\end{filecontents*}
\DTLloaddb{missing}{data/rq-inspection.csv}
\DTLgetvalue{\MissPLow}{missing}{1}{\dtlcolumnindex{missing}{P}}
\DTLgetvalue{\MissPHigh}{missing}{4}{\dtlcolumnindex{missing}{P}}
\DTLgetvalue{\MissRLow}{missing}{5}{\dtlcolumnindex{missing}{R}}
\DTLgetvalue{\MissRHigh}{missing}{2}{\dtlcolumnindex{missing}{R}}
\DTLgetvalue{\MissFLow}{missing}{5}{\dtlcolumnindex{missing}{F}}
\DTLgetvalue{\MissPAvg}{missing}{6}{\dtlcolumnindex{missing}{P}}

\def\UrlBreaks{\do\A\do\B\do\C\do\D\do\E\do\F\do\G\do\H\do\I\do\J
\do\K\do\L\do\M\do\N\do\O\do\P\do\Q\do\R\do\S\do\T\do\U\do\V
\do\W\do\X\do\Y\do\Z\do\[\do\\\do\]\do\^\do\_\do\`\do\a\do\b
\do\c\do\d\do\e\do\f\do\g\do\h\do\i\do\j\do\k\do\l\do\m\do\n
\do\o\do\p\do\q\do\r\do\s\do\t\do\u\do\v\do\w\do\x\do\y\do\z
\do\.\do\@\do\\\do\/\do\!\do\_\do\|\do\;\do\>\do\]\do\)\do\,
\do\?\do\'\do+\do\=\do\#} 
\maketitle

\section{Introduction}
\label{sec:introduction}

In the financial market, each type of investment fund, such as
UCITS\footnote{UCITS (Undertakings for the Collective Investment in
  Transferable Securities funds) refers to a regulatory framework that
  allows for the sale of cross-Europe mutual funds.}, is presented to
clients through one or more financial documents (such as KIIDs --- Key
Investor Information Document --- and prospectuses).  Before these
documents are made publicly available, they are submitted to national
financial regulators, who check their compliance with the
\emph{content requirements} prescribed by relevant national and
international laws.

The concept of ``content requirement (found in the law)'' has been recently
proposed by \citet{ceci2024defining}, who define content
requirements found in the law as ``\emph{deontic rules} requiring that
some information is contained within an official document''.
In this work, we adopt the same definition, and consider 
the content
requirements of financial documents, also called \emph{financial
  content requirements}.
For example, the following article
of a national law regulating financial market~\cite{cssf2010the}
``\emph{the frequency of the calculation of issue prices (should be
	presented)}''
prescribes the inclusion of the information related to
``issue price'' in the prospectuses of UCITS funds.

After the submission of a financial document,
agents of a financial regulator peruse the document and manually
identify the passages of text (e.g., sentences or paragraphs) related
to each mandated information type to ensure the document completeness,
since missing information may lead to substantial fines and cause legal problems and severe
investment losses when conducting activities on financial markets.
However, the manual identification of information types is a non-trivial task.
A financial document is usually lengthy with typically hundreds of
pages and more than 3000 sentences, and contains several tables and lists.
Since agents have to carefully read and analyze every sentence to avoid  any misunderstanding of the content,
considerable time could be spent by simply going through the entire document.
The amount of manual work involved often leads to higher fund setup costs and longer time-to-market for investment funds.
Therefore, it is important to develop approaches for automatically identifying
the passages of text related to content requirements, which can then further
enable automated compliance checking techniques.

Mining financial data play a critical role in improving the quality of
financial services.  Existing studies focus on mining financial data
from both Web media (e.g., financial news and discussion
boards~\cite{li2017web}) and traditional financial documents (e.g.,
annual financial reports and 10-K~\cite{li2016table}).  In
contrast to these studies, we focus on the task of tracing content
requirements in financial documents, which is important to ensure their
completeness.  Although tracing requirements in
documents have been widely studied in the areas of information
extraction and software
engineering~\cite{cleland2014software,grishman2019twenty}, existing
approaches may not suit our task due
to the complexity and the domain-specific vocabulary of financial documents.
In the field of information
extraction for general-purpose documents, the work typically focuses on extracting entities and
relations (e.g., named entities such as persons and locations) from natural language (NL)
documents~\cite{li2020survey,jiang2007systematic} instead of
identifying sentences related to content requirements.
Although recent advances in deep learning
 make the accurate identification of sentences
possible~\cite{devlin2019bert,joshi2020spanbert},
the large training set required to train the underlying models~\cite{zhang2020revisiting} is usually
unavailable in the financial area, due to the cost of annotating
thousands of financial documents by domain experts and the differences
among documents determined by national regulations.
In software
engineering, several studies~\cite{cleland2010machine,torre2020ai}
infer trace links between high-level NL
requirements (e.g., regulatory code) and low-level NL requirements
(e.g., privacy policies).
However, a
typical NL requirement is often explained with one or two sentences in
the regulatory text~\cite{hayes2006advancing}; in contrast, the
meaning of the same sentence in financial documents can differ
across different contexts, which is seldom the case for
SE requirements (where ambiguities are typically avoided).  Hence, we need to
design algorithms to trace financial content requirements while fully accounting for
the characteristics of financial documents.

In this paper, we present \approach (Financial Information Type
Identification), an approach to trace
financial content requirements with multi-granularity text analysis.
Its basic idea is to learn the characteristics of sentences related to an information type combining both IR and ML techniques from a small set of labeled documents.
Given a new financial document, \approach selects sentences for an information type based on the analysis results of IR and ML models.
Specifically, \approach first preprocesses financial documents with typical natural language processing (NLP) techniques.
To conduct efficient analysis on thousands of sentences in a new financial document, a set of candidate sentences is retrieved by comparing the similarity between the related sentences in the labeled documents and every sentence in the new document.
For the candidate sentences, \approach conducts a fine-grained analysis with IR and ML techniques.
To capture the meaning of different sentences
\approach uses similarity-based analysis with IR techniques to compare the words, sentences, and contexts between candidate sentences and the sentences related to an information type in the labeled documents.
In addition, we also mine and learn a set of features relevant to an information type with feature-based analysis and train ML-based statistical models.
According to the similarity- and feature-based analysis, \approach ranks each new sentence.
At last, \approach uses a heuristic-based selector to select the final sentences.
We built a list of domain-specific phrases that are commonly used to explain an information type (e.g., financial jargon), as well as some excluded synonyms which are seldom used to express that information type according to domain experts' suggestions and the labeled documents.
By considering both sentence ranking and phrase lists,
\approach identifies sentences for an information type from the candidate sentences.

We evaluated \approach using the content requirements for UCITS prospectuses.
Three domain experts manually annotated sentences related to five representative information types for 100 UCITS prospectuses to form a dataset.
Experimental results show \approach can accurately identify the
sentences for the five information types with average precision and recall values of {\FITIPAvg} and {\FITIRAvg}, respectively; it significantly outperforms the baselines based on keywords and language models.
Further, \approach can help regulator's agents detect about 80\% of missing information types.
Last, \approach is effective even when the number of labeled documents
is limited. With more than 40 labeled documents, the precision value of \approach is still higher than 70\% for identifying most information types.

To summarize, the main contributions of this paper are:
\begin{itemize}
	\item the first work, to the best of our knowledge, on
          tracing content requirements in financial documents, which
          is important for financial enterprises and regulators to
          further enable automated compliance techniques;
	\item the \approach approach, which addresses the problem of
	automated information type identification: it combines IR and ML to conduct fine-grained analysis on the sentences related to each information type;
	\item an extensive evaluation on the effectiveness of \approach.
\end{itemize}

The rest of the paper is organized as follows. Section~\ref{sec:background}
explains the characteristics of financial documents and their content requirements.
Section~\ref{sec:algorithm} describes the core
algorithms of \approach. Section~\ref{sec:evaluation} reports on the evaluation
of \approach. Section~\ref{sec:disc-pract-impl} discusses practical
implications. Section~\ref{sec:related_work} surveys related work. Section~\ref{sec:conclusion}
concludes the paper and provides directions for future work.

\section{Background}
\label{sec:background}

\subsection{Financial Documents}
\label{sec:financial-doc}

In the financial domain, every investment fund is required to provide informative financial documents.
These documents help the financial regulator and fund clients understand all relevant and critical information of an investment.
For example,  KIIDs describe the nature and key risks of the fund, while a prospectus provides details about an investment offering to the public.
Such financial documents mainly use natural language together
with auxiliary tables and mathematical formulae.
These documents are the key instruments to guarantee the compliance and controllability of an investment.

\figurename~\ref{fig:snippet} shows a snippet of financial document from a prospectus.
This snippet explains two types of required information in a prospectus, including \textit{calculation method for issue price} and \textit{issue conditions and procedures}.
For example, the following sentences ``The swing factor
	may normally not exceed 3\% of the net asset value of a sub-fund
	\dots In such case, affected shareholders shall be informed as soon
	as reasonably practicable thereafter \dots'' refer to the
\emph{calculation method for issue price}.

Financial documents for investment funds have several key characteristics.
First, they are lengthy with typically hundreds of pages. To minimize investment risks, financial documents must provide certain elements of information required by the regulator body for any investment fund (and its sub-funds). Based on 100 randomly selected UCITS prospectuses, our statistics show that a prospectus has on average 119 pages with around 3000 sentences.
Our statistics further show that an information type is usually explained with 3 to 36 sentences, on average, depending on the way an information type is explained by the investment company.
These characteristics make the document difficult to read and thoroughly analyze.

\subsection{Content Requirements}

\begin{figure}[tb]
	\centering
	
\tikzstyle{round_rect} = [rectangle, rounded corners, minimum width=1.5cm, minimum height=0.5 cm,text centered, draw=black, fill=white!30]
\tikzstyle{rect} = [rectangle, minimum width=1.5cm, minimum height=0.8cm, text centered, text width=1.5cm, draw=black, fill=white!30]
\tikzstyle{text_node} = [rectangle, minimum width=0.5cm, minimum height=0.35cm, text centered, text width=4cm, draw=white, fill= white!30]

\resizebox{0.48\textwidth}{!}{
	\begin{tikzpicture} [node distance=5mm, >=latex]

\node (page)[inner sep=0pt]
		{\includegraphics[width=.13\textwidth]{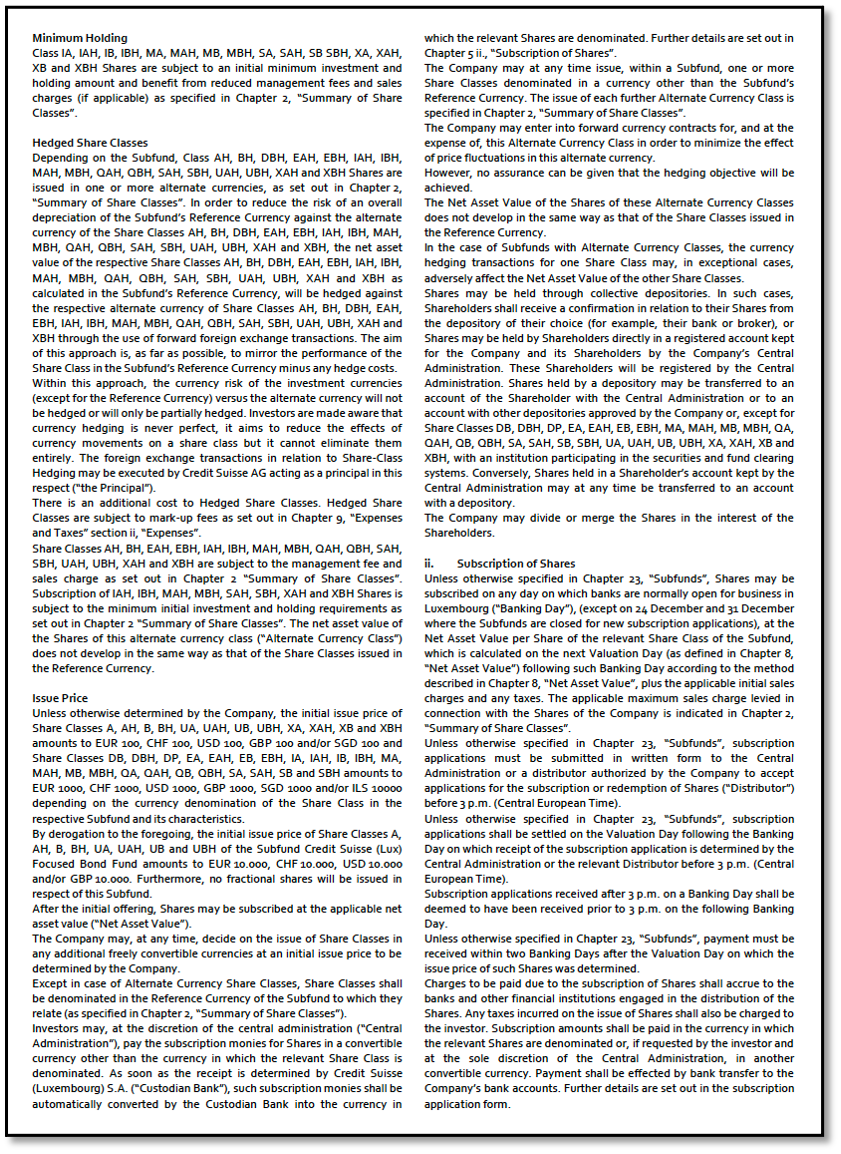}};
		\node (locate) [draw=blue, minimum width=1.0cm, minimum height=1.1cm, line width=0.2mm, right = of page, xshift=-1.5cm, yshift=-0.4cm] {};
		\node (ptext)[text_node, minimum width=0.2cm, below= of page, yshift=0.5cm] {A page of financial document};

		\draw[fill=blue!10, ] (1.0,0.2) -- (1.0,-1) -- (1.4,-1.7) -- (1.4,3.6)
		-- cycle;

\node (snippet)[inner sep=0pt, right=of page, yshift= 1cm]
		{\includegraphics[width=.5\textwidth]{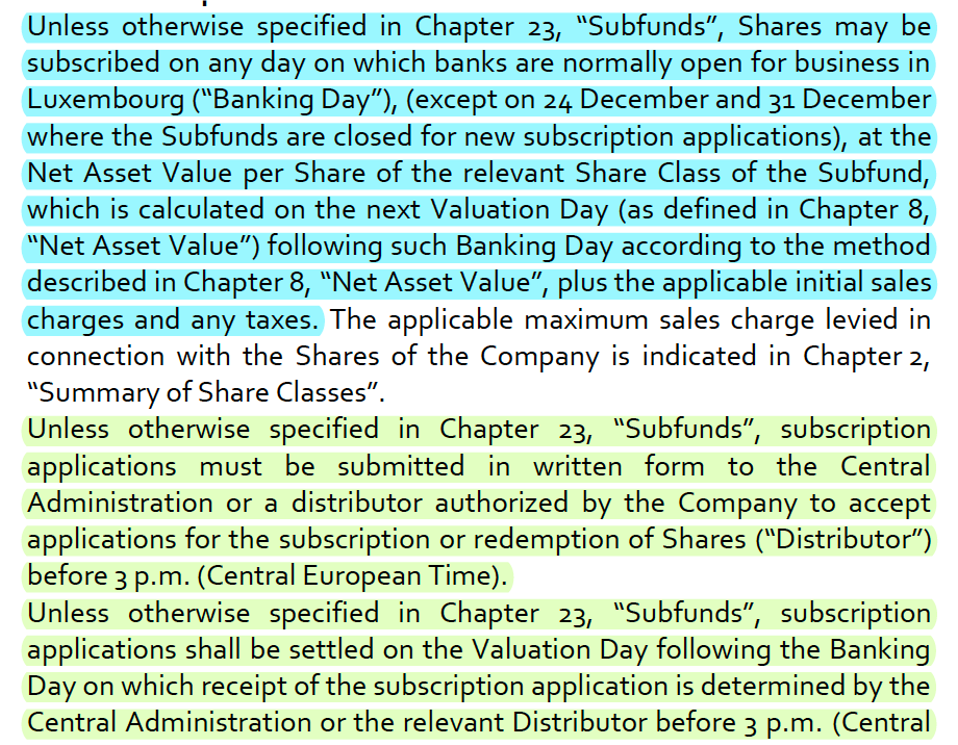}};
		\node (method) [text_node, minimum width=0.3cm, above= of snippet, yshift=-0.2cm, xshift=-2cm] {Calculation method for issue price};
		\node (issue) [text_node, minimum width=0.3cm, above= of snippet, yshift=-0.2cm, xshift=2cm] {Issue conditions and procedures};
		\node (point1)[draw=white, below = of method, yshift=0.2cm]{};
		\node (point2)[draw=white, below = of issue, yshift=-3.6cm]{};
		\draw [arrow, ->, thick] (method)--(point1);
		\draw [arrow, ->, thick] (issue)--(point2);

	\end{tikzpicture}
}
 	\caption{A Snippet of Financial Document}
	\label{fig:snippet}
\end{figure}

Before presenting them to the public,
financial documents should be inspected by the regulator.
One of the main inspection tasks is to check the completeness of the documents.
In practice, the regulator defines content requirements for each type of financial document, specifying types of information that should be present within a document.
For example, for a UCITS prospectus submitted to a national regulator,
regulatory requirements stipulate about 124 information types, including for example the description of the calculation method for issue price and issue conditions and procedures (as presented in Fig.~\ref{fig:snippet}).
The regulator must analyze the prospectus to identify the sentences discussing each information type specified by content requirements.
Since missing information may cause legal problems and severe investment losses,
ensuring content completeness is usually the first and most fundamental procedure before looking into the details of the financial documents.
However, due to the length and peculiarities of the text in financial documents, this is usually a challenging task.

\section{Information Type Identification}
\label{sec:algorithm}

\begin{algorithm}[tb]
	\newcommand\mycommfont[1]{\footnotesize\ttfamily\textcolor{blue}{#1}}
	\SetCommentSty{mycommfont}
	\small
	\caption{\approach}
	\label{alg:financial-doc-checking}
	\KwIn{Financial document $d$ for analysis \newline
		Information type $t$ \newline
		A set of labeled documents $D^L$, with $d \not \in D^L$\newline
		Number of candidate sentences for analysis $n_c$\newline
		Related phrase list $\mathit{rlist}$ and unrelated phrase list $\mathit{ulist}$ \newline
Threshold $\theta$ for highly similar sentences
	}
	\KwOut{Sentences $S$ in $d$ related to $t$}
	\tcp{Step 1: Pre-processing}
	$D^L, d \gets$ \emph{preProcessing}$(D^L, d)$\; \label{line:preprocessing}
	\tcp{Step 2: Identify candidate sentences}
	$S^L_{r_t}\gets$\emph{getRelatedSentences}$(D^L, t)$\;  \label{line:candidates-s}
	$S^L_{u_t}\gets$\emph{getUnrelatedSentences}$(D^L, t)$\;
	Candidates $S^d_{c_t}\gets$\emph{candidateSelection}$(d, S^L_{r_t}, n_c)$\; \label{line:candidates-e}
	\tcp{Step 3: Fine-grained sentence analysis}
	\ForEach{$s\in S^d_{c_t}$}{ \label{line:fine-grained-s}
			$s.\mathit{group}, s.\mathit{avg\_s}, s.\mathit{max\_s}, s.\mathit{word}$ $\gets$\emph{calcIRSimilarity}$(s, S^L_{r_t}, S^L_{u_t})$\;
	} \label{line:ir-e}
	Instances $I_\mathit{pos}, I_\mathit{neg} \gets$\emph{getInstances}$(S^L_{r_t}, S^L_{u_t})$\;\label{line:ml-s}
	Classifier $\mathit{cls}\gets$\emph{trainRandomForest}$(I_\mathit{pos}, I_\mathit{neg})$\;
	\ForEach{$s\in S^d_{c_t}$}{
		$s.\mathit{prob}\gets$\emph{calcMLProbability}$(cls, s)$\;
	} \label{line:fine-grained-e}
	\tcp{Step 4: Sentence selection}
	$n_r\gets$ \emph{calcAvgNumOfRelatedSentences}$(S^L_{r_t})$\;
	$S\gets$ \emph{Selector}$(S^d_{c_t}, n_r, rlist, ulist, \theta)$\; \label{line:final-sentences}
	return $S$\;
\end{algorithm}

In this section, we present our algorithm (\approach: Financial Information Type Identification) to automatically identify the sentences related to an information type in financial documents.
Its pseudocode is shown in Algorithm~\ref{alg:financial-doc-checking}.
\approach takes as input an unlabeled document $d$ for analysis, the information
type $t$ to identify (from the content requirements), a set of labeled
documents $D^L$ (in which sentences related to $t$ have been annotated
by domain experts) with $d \not \in D^L$, and some auxiliary parameters; it returns a set of
sentences $S$ from $d$ related to $t$.

\approach has four main steps:
pre-processing (\S~\ref{sec:preprocessing}),
candidate sentence identification (\S~\ref{sec:candidates}),
fine-grained sentence analysis (\S~\ref{sec:fine-grained}), and
sentence selection (\S~\ref{sec:sentence-sel}).

It first pre-processes (line~\ref{line:preprocessing}) the document
with a standard NLP pipeline (including sentence splitting,
tokenization, stop words removal, stemming, named entity recognition). Then, for an information type $t$, \approach identifies a set of
candidate sentences in $d$ for fine-grained analysis
(lines~\ref{line:candidates-s}--\ref{line:candidates-e}). Next,
\approach analyzes the candidate sentences with information retrieval
(IR) and machine learning (ML); it assigns scores to each candidate
sentence
(lines~\ref{line:fine-grained-s}--\ref{line:fine-grained-e}). Last, a
heuristic-based selector is applied to select the final sentences that
are most likely related to $t$ (line~\ref{line:final-sentences}).

\subsection{Pre-processing}
\label{sec:preprocessing}

\begin{figure*}[tb]
	\centering
	
\tikzstyle{round_rect} = [rectangle, rounded corners, minimum width=1.5cm, minimum height=0.5 cm,text centered, draw=black, fill=white!30]
\tikzstyle{rect} = [rectangle, minimum width=1.5cm,  text centered, text width=1.5cm, draw=black, fill=white!30]
\tikzstyle{text_node} = [rectangle, minimum width=0.5cm, text centered, text width=4cm, draw=white, fill= white!30]
\tikzstyle{text_model} = [rectangle, minimum width=0.8cm, text centered, text width=0.8cm, draw=white, fill= white!30]

\tikzstyle{list_box} = [rect,align=left ,minimum width=2.39cm,text width=3.19cm, minimum height= 0.5cm]

\resizebox{1\textwidth}{!}{
	\begin{tikzpicture} [node distance=0.3mm, >=latex]
	
	\node (doc) [rect, draw=none, text width=3cm]{ \includegraphics[width=\textwidth]{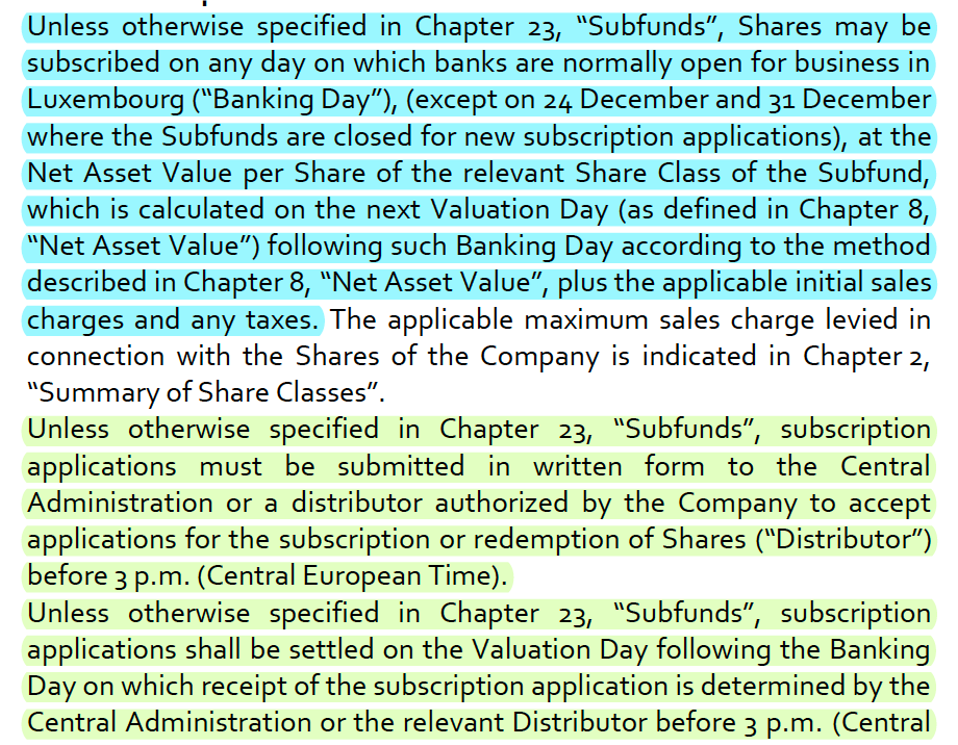}};
	\node (doc_text) [rect, draw=none, below = of doc, text width=4cm] {Unlabeled document $d$};
	
	\node (s1) [text_node, right= = of doc, minimum width=0.3cm, text width=0.3cm, xshift=2cm, yshift=1cm]{$s_1$:};
	\node (s1_text) [text_node, right= = of s1, text
	width=7.3cm, xshift=2.5cm, yshift=1cm, align=left]{\dots\emph{subfund close new subscript application}\dots};
	
	\node (s2) [text_node, right= = of doc, minimum width=0.3cm, text width=0.3cm, xshift=2cm, yshift=0cm]{$s_2$:};
	\node (s2_text) [text_node, right= = of s2, text
	width=7.3cm, xshift=2.5cm, yshift=0cm, align=left]{\dots\emph{applicable maximum sale charge levy}\dots};
	
	\node (s3) [text_node, right= = of doc, minimum width=0.3cm, text width=0.3cm, xshift=2cm, yshift=-1cm]{$s_3$:};
	\node (s3_text) [text_node, right= = of s3, text
	width=7.3cm, xshift=2.5cm, yshift=-1cm, align=left]{\dots\emph{subscript application submit written form}\dots};
			
	\node (s0) [text_node, draw=none, above= of s1_text, text
	width=7.3cm]{$\cdots\cdots$};
	
	\node (sn) [text_node, draw=none, below= of s3_text, text
	width=7.3cm]{$\cdots\cdots$};

	\draw [arrow,->] (doc) -- (s1);
	\draw [arrow,->] (doc) -- (s2);
	\draw [arrow,->] (doc) -- (s3);
	
	\node (s1_v) [rect, right= of s1_text, text
	width=5.3cm, xshift=0.7cm]{\dots 4.2, 2.6, 0.0, 8.0, \dots, 3.0, 0.0, 5.0 \dots};
	
	\node (s2_v) [rect, right= of s2_text, text
	width=5.3cm, xshift=0.7cm]{\dots 6.3, 2.6, 1.0, 8.0, \dots, 3.0, 2.2, 7.5 \dots};
	
	\node (s3_v) [rect, right= of s3_text, text
	width=5.3cm, xshift=0.7cm]{\dots 2.1, 0.0, 1.0, 8.0, \dots, 0.0, 2.2, 2.5 \dots};
	
	\node (s_v_text) [rect, draw=none, below = of s3_v, text width=5.3cm] {The context vector of each sentence using TF-IDF measurement};
		
	\node (sg_v) [rect, above= of s1_v, text
	width=5.3cm, yshift=1.49cm]{\dots 8.4, 2.6, 3.0, 0.0, \dots, 1.5, 8.8, 7.5 \dots};
	
	\node (voc1) [text_node, above= of sg_v,minimum width=0.3cm, text width=0.3cm, xshift=-2.4cm, yshift=0.3cm, rotate around={45:(0,0)}, ]{$\cdots$};
	\node (voc2) [text_node, right= of voc1,minimum width=0.3cm, text width=0.3cm, rotate around={45:(0.55,0)}]{application};
	\node (voc3) [text_node, right= of voc2,minimum width=0.3cm, text width=0.3cm, rotate around={45:(0.55,0)}]{close};
	\node (voc4) [text_node, right= of voc3,minimum width=0.3cm, text width=0.3cm, rotate around={45:(0.55,0)}]{form};
	\node (voc5) [text_node, right= of voc4,minimum width=0.3cm, text width=0.3cm, rotate around={45:(0.55,0)}]{levy};
	\node (voc6) [text_node, right= of voc5,minimum width=0.3cm, text width=0.3cm, rotate around={45:(0.55,0)}]{$\cdots$};
	\node (voc7) [text_node, right= of voc6,minimum width=0.3cm, text width=0.3cm, rotate around={45:(0.55,0)}]{subfund};
	\node (voc8) [text_node, right= of voc7,minimum width=0.3cm, text width=0.3cm, rotate around={45:(0.55,0)}]{submit};
	\node (voc9) [text_node, right= of voc8,minimum width=0.3cm, text width=0.3cm, rotate around={45:(0.55,0)}]{subscript};
	\node (voc10) [text_node, right= of voc9,minimum width=0.3cm, text width=0.3cm, rotate around={45:(0.55,0)}]{$\cdots$};
	
	\node (voc_text) [text_node, above = of sg_v, text width=1.3cm, yshift=1.5cm] {\textbf{Vocabulary}};

	\node (point_s1_text) [text_node, minimum width=0.0cm, text width=0.0cm, right = of s1_text, xshift=-0.6cm] {};
	\node (point_s2_text) [text_node, minimum width=0.0cm, text width=0.0cm, right = of s2_text, xshift=-0.6cm] {};
	\node (point_s3_text) [text_node, minimum width=0.0cm, text width=0.0cm, right = of s3_text, xshift=-0.6cm] {};
	\node (point_s1_v) [text_node, minimum width=0.0cm, text width=0.0cm, left = of s1_v, xshift=0.0cm] {};
	\node (point_s2_v) [text_node, minimum width=0.0cm, text width=0.0cm, left = of s2_v, xshift=0.0cm] {};
	\node (point_s3_v) [text_node, minimum width=0.0cm, text width=0.0cm, left = of s3_v, xshift=0.0cm] {};
	\node (point_s0_text) [text_node, minimum width=0.0cm, text width=0.0cm, above = of point_s1_text, yshift=0.2cm] {};
	\node (point_sn_text) [text_node, minimum width=0.0cm, text width=0.0cm, below = of point_s3_text, yshift=-0.2cm] {};
	\draw [arrow,->] (point_s1_text) -- (point_s1_v);
	\draw [arrow,->] (point_s2_text) -- (point_s2_v);
	\draw [arrow,->] (point_s3_text) -- (point_s3_v);
	\draw [arrow,->, dashed] (point_s1_text) -- (point_s2_v);
	\draw [arrow,->, dashed] (point_s2_text) -- (point_s1_v);
	\draw [arrow,->, dashed] (point_s2_text) -- (point_s3_v);
	\draw [arrow,->, dashed] (point_s3_text) -- (point_s2_v);
	\draw [arrow,->, dashed] (point_s0_text) -- (point_s1_v);
	\draw [arrow,->, dashed] (point_sn_text) -- (point_s3_v);
	
	\node (labeled) [rect, above= = of doc, text
	width=2.5cm, yshift=3.12cm]{$s^L_{1r_t}, \dots, s^L_{ir_t}$};
	\node (labeled_text) [rect, draw=none, below = of labeled, text width=5.3cm] {Related sentences $S^L_{r_t}$};
	\node (point_s_labeled) [text_node, minimum width=0.0cm, text width=0.0cm, left = of sg_v, xshift=0.0cm] {};
	\draw [arrow,->] (labeled) -- node [anchor=south] {Group vector}(point_s_labeled);
	
	\draw [arrow,<->] (sg_v) -- (s1_v);
	\node (point_cosine) [text_node, minimum width=0.0cm, text width=0.0cm, above = of s1_v, yshift=0.6cm, xshift=-0.14cm] {};
	
	\node (s_set) [rect, right= of s1_v, text
	width=2.0cm, yshift=1.22cm, xshift=0.5cm]{$s_1, s_2, s_{10}, \dots$};
	\node (s_set_text) [rect, draw=none, below = of s_set, text width=2.3cm] {Candidate sentences};
	\draw [arrow,->] (point_cosine) -- node [anchor=south] {Cosine similarity}(s_set);

	\node (big_box)[draw=black, fit={(s0)(s1)(s1_text)(s2)(s2_text)(s3)(s3_text)(sn)},minimum width=1.4cm,text width=7.3cm, minimum height=2.5cm, xshift=-0.2cm]{};
	
	\node (s_process) [rect, draw=none, below = of big_box, text width=5.3cm] {Preprocessed sentences};
	
	\end{tikzpicture}
      }

 	\caption{Workflow of pre-processing and candidate sentence identification}
	\label{fig:candidates-flow}
\end{figure*}

Financial documents are typically available in PDF format. To ease their
manipulation, we convert them to a plain-text format using an
off-the-shelf converter PDFBox~\cite{pdfbox}.
We then apply a standard NLP pipeline to preprocess the text.
The text is first split into sentences with Stanford CoreNLP~\cite{stanfordnlp}.
Then, tokenization is applied to identify the words in a sentence.
We remove the stopwords~\cite{stopwords} and convert each word into its root form with the Porter stemming algorithm~\cite{stemming}.
In addition, the content of financial documents often includes named entities such
as numbers, person names, dates, and web addresses. To leverage the knowledge of these named entities, we perform named entity
recognition~\cite{stanfordnlp} on the input document to generalize
these named entities with their category names.
Once the above steps are completed, we obtain a list of preprocessed, simplified words and sentences from the financial documents.

\subsubsection*{Application to the running example}
As shown in \figurename~\ref{fig:candidates-flow}, given an unlabeled document $d$,
\approach transforms it into a list of preprocessed sentences $s_1, s_2, s_3, \cdots, s_i$.
For example, a sentence ``Annex I takes effect from 1
January 2016'' becomes ``Annex NUMBER take effect DATE''.
After preprocessing, the number ``I'' and date ``1 January 2016'' are transformed into ``NUMBER'' and ``DATE'', respectively.

\subsection{Candidate Sentence Identification}
\label{sec:candidates}
Although financial documents include thousands of sentences, a specific information type is usually addressed by less than 50 sentences.
Conducting fine-grained analysis on the entire document may therefore be
impractical on commodity hardware.
To solve this problem, \approach tries to efficiently filter the majority of unrelated sentences in the pre-processed document $d$ and identify a small number of candidate sentences for further analysis.
The basic hypothesis for candidate sentence identification is that sentences in $d$ that are similar to existing sentences related to $t$ in the labeled documents $D^L$ may also be related to $t$.
Therefore, we calculate similarity between sentences in $d$ and sentences annotated as related to $t$ in $D^L$. We take the top-$n_c$ most similar sentences as candidates for fine-grained analysis.

Specifically, we collect the sentences annotated as related to $t$ in $D^L$.
We transform this group of sentences into a single vector (denoted as ``group vector'') using a standard IR model: the bag-of-words model~\cite{manning2008introduction}.
Given a corpus (e.g., documents in $D^L$), the bag-of-words model gets its vocabulary (i.e., all non-duplicated words) and represents a piece of text into a vector, where the length of the vector is equal to the size of the vocabulary.
In our context, each dimension of the group vector means a word in the vocabulary. If the group of related sentences does not contain a word, the value of the corresponding dimension is 0; otherwise, the value is computed by the TF-IDF (Term Frequency-Inverse Document Frequency)~\cite{manning2008introduction} of the word.
TF-IDF is defined as:
\begin{equation}\label{eq:tf-idf}
\text{TF-IDF}_{w, \mathit{text}}=f_{w, \mathit{text}}\times \log \frac{N}{n_w},
\end{equation}
where $f_{w, \mathit{text}}$ denotes the number of times that $w$ occurs in $\mathit{text}$ (e.g., the group of related sentences),
$N$ is the number of sentences in the corpus,
and $n_w$ is the number of sentences in the corpus that contain $w$.
TF-IDF based vectors assume that sentences can be represented with the frequently used and informative words, where TF ($f_{w, \mathit{text}}$) calculates the frequency of words and IDF ($\log \frac{N}{n_w}$) identifies informative words that are not used in almost all the sentences. For example, TF-IDF can identify words such as  ``swing'' (as in ``swing factor'') and ``dilution'' (as in `` dilution adjustment'') in the example sentences in Section~\ref{sec:financial-doc} as informative words,
since they are commonly expected in financial documents but only used in specific contexts.

In this step, we do not use more complex vectorization models (e.g., deep
learning based sentence embedding), as the bag-of-words model is
easy to deploy and understand, it does not require a large
domain-specific corpus (UCITS prospectuses in our case) to learn the embedding of domain-specific words and phrases, which is usually not available.

Further, we transform each sentence in the document $d$ into a TF-IDF based vector as follows.
For each sentence, we collect its surrounding $n_\mathit{cxt}$ sentences,
which represent the context of the current sentence.
For example, if $n_\mathit{cxt}=1$, the context of a sentence includes its previous sentence and the next sentence.
We compute the frequency (i.e., TF) and the inverted document frequency (i.e., IDF) of each term in the sentence itself and its context.
We transform all these TF-IDF based values into a single vector (hereafter called ``context vector'').
We consider the context of a sentence because a single sentence may not contain enough information for our analysis.
The context of a sentence provides valuable information for understanding it.

Last, we compute the similarity between the group vector and the
context vector of each sentence in $d$ using cosine
similarity~\cite{manning2008introduction}, defined as:
\begin{equation}\label{eq:cosine}
\mathit{sim}(\vec{v_1}, \vec{v_2})=\frac{\vec{v_1} \vec{v_2}}{\left| \vec{v_1} \right| \left| \vec{v_2} \right|},
\end{equation}
where $\vec{v_1} \vec{v_2}$ is the inner product of the two vectors and $\left| \vec{v_1} \right| \left| \vec{v_2} \right|$ is the product of the 2-norm for these vectors.
We then rank and select $n_c$ sentences in $d$ as candidates for further analysis.

In this study, the number of surrounding sentences $n_\mathit{cxt}$ is set to 1
and the number of candidate sentences $n_c$ is set to 200.
Our preliminary experiment shows that on average the candidate sentences contain 90\% of sentences related to an information type, therefore effectively filtering 94.2\% unrelated sentences and preserve the vast majority of related sentences.

\subsubsection*{Application to the running example}
The workflow of candidate sentence identification is presented in \figurename~\ref{fig:candidates-flow}.
On the top of the figure, \approach transforms all sentences in $S^L_{r_t}$, which are annotated as related to $t$ in $D^L$, into a group vector.
Each dimension of the vector represents a word in vocabulary, such as ``application'', ``close'', and ``form'' in the example.
Meanwhile, as shown at the bottom of the figure, \approach transforms each preprocessed sentence in $d$ into a context vector.
For example, when constructing the context vector for $s_2$,
\approach computes the TF-IDF values of words in $s_2$, as well as in its context $s_1$ and $s_3$.
All these TF-IDF values are transformed into a single vector $[\dots 6.3, 2.6, 1.0, 8.0, \dots, 3.0, 2.2, 7.5 \dots]$.
At last, \approach computes the cosine similarity between the vector for $S^L_{r_t}$ and each vector of the sentences in $d$ to select candidate sentences.

\subsection{Fine-grained Sentence Analysis}
\label{sec:fine-grained}

\begin{figure*}[tb]
	\centering
	
\tikzstyle{round_rect} = [rectangle, rounded corners, minimum width=1.5cm, minimum height=0.5 cm,text centered, draw=black, fill=white!30]
\tikzstyle{rect} = [rectangle, minimum width=1.5cm,  text centered, text width=1.5cm, draw=black, fill=white!30]
\tikzstyle{text_node} = [rectangle, minimum width=0.5cm, text centered, text width=4cm, draw=white, fill= white!30]
\tikzstyle{text_model} = [rectangle, minimum width=0.8cm, text centered, text width=0.8cm, draw=white, fill= white!30]

\tikzstyle{list_box} = [rect,align=left ,minimum width=2.39cm,text width=3.19cm, minimum height= 0.5cm]

\resizebox{1\textwidth}{!}{
	\begin{tikzpicture} [node distance=0.3mm, >=latex]
	
	\node (labeled1) [rect, text
	width=2.0cm]{$s^L_{1r_t}, \dots, s^L_{ir_t}$};
	\node (labeled_text1) [rect, draw=none, below = of labeled1, text width=3.3cm] {Related sentences $S^L_{r_t}$};
	
	\node (vec_group) [rect, right = of labeled1, minimum width=1cm, text
	width=1.8cm, xshift=0.7cm]{group\\vector};
				
	\node (similar1) [round_rect, right = of vec_group, minimum width=1cm, text
	width=1.7cm, xshift=0.7cm]{cosine similarity};
	
	\node (vec_context_1) [rect, right = of similar1, minimum width=1cm, text
	width=1.5cm, xshift=0.7cm]{context vector};
	\node (candidate_text) [rect, draw=none, below = of vec_context_1, text width=2.5cm] {Candidate sentence $s_i$};
	
	\draw [arrow,->] (labeled1) -- (vec_group);
	\draw [arrow, ->] (vec_group) -- (similar1);
	\draw [arrow, ->] (vec_context_1) -- (similar1);
	
	\node (sim_sg) [text_node, below = of similar1, minimum width=1.5cm, text width=1.5cm, yshift=-1.0cm]{\color{blue}{$s.\mathit{group}$}};
	\draw [arrow,->] (similar1) -- (sim_sg);
	\node (example1) [rect, draw=none, below = of sim_sg, text width=3.3cm, xshift=-1.5cm, yshift=-0.2cm] {(A) Group similarity};
	
\node (labeled2) [rect, right = of vec_context_1, text
	width=2.0cm, xshift=1cm]{$s^L_{1r_t}, \dots, s^L_{ir_t}$};
	\node (labeled_text2) [rect, draw=none, below = of labeled2, text width=3.3cm] {Related sentences $S^L_{r_t}$};
	
	\node (vec_si) [rect, right = of labeled2, minimum width=1cm, text
	width=3.3cm, xshift=0.7cm]{context vector of $s^L_{ir_t}$};
	\node (vec_sn) [text_node, draw=none, above= of vec_si, text width=3cm]{$\cdots\cdots$};
	\node (vec_s1) [rect, above = of vec_sn, minimum width=1cm, text
	width=3.3cm]{context vector of $s^L_{1r_t}$};
	
	\node (similar2) [round_rect, right = of vec_si, minimum width=1cm, text
	width=1.7cm, xshift=0.7cm]{cosine similarity};
	
	\node (vec_context_2) [rect, right = of similar2, minimum width=1cm, text
	width=1.5cm, xshift=0.7cm]{context vector};
	\node (candidate_text) [rect, draw=none, below = of vec_context_2, text width=2.5cm] {Candidate sentence $s_i$};
	
	\draw [arrow,->] (labeled2.east) -- (vec_s1.west);
	\draw [arrow,->] (labeled2) -- (vec_si);
	\draw [arrow,->] (vec_si) -- (similar2);
	\draw [arrow,->] (vec_s1.east) -- (similar2.west);
	\draw [arrow,->] (vec_context_2) -- (similar2);
	
	\node (sim_ss) [text_node, below = of similar2, minimum width=1.5cm, text
	width=1.5cm, yshift=-1.0cm]{\color{blue}{$s.\mathit{avg\_s}$, $s.\mathit{max\_s}$}};
	\draw [arrow,->] (similar2) -- (sim_ss);
	\node (example2) [rect, draw=none, right = of example1, text width=3.5cm, xshift=7.5cm] {(B) Sentence similarity};
\node (labeled3) [rect, below = of labeled1, text
width=2.0cm, yshift=-4cm]{$s^L_{1r_t}, \dots, s^L_{ir_t}$};
\node (labeled_text3) [rect, draw=none, below = of labeled3, text width=3.3cm] {Related sentences $S^L_{r_t}$};

\node (vec_imp) [rect, right = of labeled3, minimum width=1cm, text
width=1.8cm, xshift=0.7cm]{$\mathit{imp}(w)$ of words};

\node (similar3) [round_rect, right = of vec_imp, minimum width=1cm, text
width=1.7cm, xshift=0.7cm]{importance similarity};

\node (vec_context_3) [rect, right = of similar3, minimum width=1cm, text
width=1.5cm, xshift=0.7cm]{words in $s_i$};
\node (candidate_text) [rect, draw=none, below = of vec_context_3, text width=2.5cm] {Candidate sentence $s_i$};

\draw [arrow,->] (labeled3) -- (vec_imp);
\draw [arrow,->] (vec_imp) -- (similar3);
\draw [arrow,->] (vec_context_3) -- (similar3);

\node (sim_imp) [text_node, below = of similar3, minimum width=1.5cm, text
width=1.5cm, yshift=-1.0cm]{\color{blue}{$s.\mathit{word}$}};
\draw [arrow,->] (similar3) -- (sim_imp);

\node (example3) [rect, draw=none, below = of example1, text width=5cm, yshift=-5.2cm] {(C) Word importance similarity};

\node (labeled4) [rect, below = of labeled2, text
width=2.1cm, yshift=-4cm]{$s^L_{1r_t}, \dots, s^L_{ir_t}$};
\node (labeled_text4) [rect, draw=none, below = of labeled4, text width=3.3cm] {Related sentences $S^L_{r_t}$};

\node (vec_posi) [rect, right = of labeled4, minimum width=1cm, text
width=3.3cm, xshift=0.7cm]{feature vector of $s^L_{ir_t}$};
\node (vec_posn) [text_node, draw=none, above= of vec_posi, text width=3cm]{$\cdots\cdots$};
\node (vec_pos1) [rect, above = of vec_posn, minimum width=1cm, text
width=3.3cm]{feature vector of $s^L_{1r_t}$};

\node (similar4) [round_rect, right = of vec_posi, minimum width=1cm, text
width=1.7cm, xshift=0.7cm]{Random forest};

\node (vec_context_4) [rect, right = of similar4, minimum width=1cm, text
width=1.5cm, xshift=0.7cm]{feature vector};
\node (candidate_text) [rect, draw=none, below = of vec_context_4, text width=2.5cm] {Candidate sentence $s_i$};

\draw [arrow,->] (labeled4) -- (vec_posi);
\draw [arrow,->] (vec_posi) -- (similar4);
\draw [arrow,->] (labeled4.east) -- (vec_pos1.west);
\draw [arrow,->] (vec_pos1.east) -- (similar4.west);

\node (labeled4_neg) [rect, below = of labeled4, text
width=2.1cm, yshift=-1.5cm]{$s^L_{1u_t}, \dots, s^L_{iu_t}$};
\node (labeled_text4_neg) [rect, draw=none, below = of labeled4_neg, text width=3.7cm] {Unrelated sentences $S^L_{u_t}$};

\node (vec_negi) [rect, right = of labeled4_neg, minimum width=1cm, text
width=3.3cm, xshift=0.7cm]{feature vector of $s^L_{iu_t}$};
\node (vec_negn) [text_node, draw=none, above= of vec_negi, text width=3cm]{$\cdots\cdots$};
\node (vec_neg1) [rect, above = of vec_negn, minimum width=1cm, text
width=3.3cm]{feature vector of $s^L_{1u_t}$};

\draw [arrow,->] (labeled4_neg) -- (vec_negi);
\draw [arrow,->] (vec_negi.east) -- (similar4.west);
\draw [arrow,->] (labeled4_neg.east) -- (vec_neg1.west);
\draw [arrow,->] (vec_neg1.east) -- (similar4.west);
\draw [arrow,->] (vec_context_4) -- (similar4);

\node (sim_pos) [text_node, below = of similar4, minimum width=1.5cm, text
width=1.5cm, yshift=-1.0cm]{\color{blue}{$s.\mathit{ml}$}};
\draw [arrow,->] (similar4) -- (sim_pos);

\node (example4) [rect, draw=none, below = of example2, text width=5cm, yshift=-5.2cm] {(D) Feature-based probability};

	\end{tikzpicture}
}
 	\caption{Workflow of fine-grained sentence analysis}
	\label{fig:analysis-flow}
\end{figure*}

\approach analyzes the relevance of candidate sentences in $d$ for an information type $t$ (denoted as $S^d_{c_t}$) with different techniques, including both similarity-based analysis and feature-based analysis.
Similarity-based analysis uses information retrieval (IR) to calculate the text similarity between a sentence in $S^d_{c_t}$ and the sentences annotated as related to $t$ in $D^L$ (denoted as $S^L_{r_t}$).
It assumes that if a sentence is similar to the existing related sentences in $S^L_{r_t}$,
this sentence is more likely to be related to $t$.
Similarity-based analysis outputs similarity values for a sentence in $S^d_{c_t}$, which indicate the degree of text similarity of the candidate sentence with the sentences in $S^L_{r_t}$.
Feature-based analysis uses machine learning (ML) to mine a set of measurable properties of sentences (features) that can distinguish related sentences from unrelated ones.
It represents each sentence with a feature vector, where each component is a feature.
Feature-based analysis uses the feature vectors of related and unrelated sentences in $D^L$ to train a statistical model.
For a sentence in $S^d_{c_t}$, the trained model outputs a probability value, which indicates the probability that the sentence is related to $t$.

Similarity- and feature-based analysis techniques analyze sentences from different perspectives: the former calculates the text similarity and the latter trains statistical models with features. They are expected to complement each other.

\subsubsection{Similarity-based analysis}
\label{sec:ir-analysis}

To perform a comprehensive comparison between sentences in $S^d_{c_t}$ and sentences in $S^L_{r_t}$,
\approach calculates similarity at different granularity levels, including group similarity, sentence similarity, and word importance similarity.
Since financial documents may use similar sentences to explain different information types (as explained in section~\ref{sec:background}), we use different granularity levels to better identify the candidate sentences that are similar with the overall context of $S^L_{r_t}$, individual sentences in $S^L_{r_t}$, and the words specified for $t$ at the same time.

\textbf{Group similarity} compares the overall similarity between $S^L_{r_t}$ and every sentence in $S^d_{c_t}$.
It is calculated as specified in Section~\ref{sec:candidates}: the group similarity is the cosine similarity between the group vector of $S^L_{r_t}$ and the context vector of a candidate sentence $s$ (denoted as $s.\mathit{group}$).

\textbf{Sentence similarity} calculates the cosine similarity between a candidate sentence $s$ and each sentence in $S^L_{r_t}$ based on their context vectors.
Given $n$ sentences in $S^L_{r_t}$,
we can get $n$ sentence-level similarity values for $s$.
Based on these values,
\approach calculates two sentence-level scores for $s$: the average and the maximum of the $n$ similarity values, denoted as $s.\mathit{avg\_s}$ and $s.\mathit{max\_s}$, respectively.
We consider $s.\mathit{max\_s}$ because, when a candidate sentence is extremely similar to some of the labeled sentences, it is very likely that it is also a related sentence.

\textbf{Word importance similarity} analyzes the importance of a word for the information type $t$ based on the labeled documents $D^L$. The score of a sentence in $S^d_{c_t}$ is calculated  according to word importance.
This is automatically determined from three aspects;
precisely, we say that the importance of a word $w$ for $t$ is determined by the extent to which it satisfies the three following conditions:
\begin{enumerate*}[label=(\alph*)]
	\item $w$ frequently appears in the related sentence set $S^L_{r_t}$;
	\item $w$ is only present in $S^L_{r_t}$;
	\item $w$ can be found in the related sentences of every labeled document.
\end{enumerate*}
The importance of a word $w$ is therefore computed as:
\begin{equation}\label{eq:importance_word}
\begin{aligned}
\mathit{imp}(w) &= \mathit{freq_{avg}}(w) \times \mathit{spec}(w) \times \mathit{univ}(w) \\
&= \frac {1}{n} \sum_{r_i \in S^L_{r_t}} \frac{\mathit{freq}(r_i, w)}{\left|r_i \right|} \times \frac {\left| S^L_{r_t}(w)\right|}{\left| S^L(w)\right|} \times \frac {\left|D^L_{r_t}(w)\right|}{\left|D^L_{r_t}\right|}
\end{aligned}
\end{equation}
The first factor in the formula calculates the average frequency of $w$ in $S^L_{r_t}$,
where $n$ is the number of sentences in $S^L_{r_t}$,
$\mathit{freq}(r_i, w)$ counts the times that $w$ appears in the $i$th related sentence $r_i$,
and $\left|r_i \right|$ is the number of words in $r_i$.
The second factor considers the specificity of $w$ to $t$,
where $\left| S^L_{r_t}(w)\right|$ is the number of sentences in $S^L_{r_t}$ that contain $w$,
and $\left| S^L(w)\right|$ is the number of sentences in $D^L$ containing $w$.
Third, we analyze the universality of $w$ in the related sentences of $D^L$,
where $\left|D^L_{r_t}(w)\right|$ is the number of documents that have related sentences containing $w$,
and $\left|D^L_{r_t}\right|$ is the number of documents having related sentences.
Intuitively, the importance of a word $w$ is determined by three conditions:
if $w$ never appears in $S^L_{r_t}$, the values of these conditions are zero; in contrast, these values are 1 if $w$ is the only word in $S^L_{r_t}$ and it never appears in other sentences.
Since all the three conditions have positive correlations with the importance of a word,
we multiply the values of the three conditions to reflect the importance of $w$.
This formula is inspired by existing work on requirement analysis in software engineering~\cite{cleland2010machine},
where they use a similar method to identify indicator terms in regulations for software requirement retrieval.

Based on $\mathit{imp}(w)$ for each word in $D^L$,
the word importance similarity for a sentence $s$ (denoted as $s.\mathit{word}$)
is calculated as $\frac {\sum_{w \in s} \mathit{imp}(w)}{\sum_{w \in S^L_{r_t}} \mathit{imp}(w)}$.
In this definition, $s.\mathit{word}$=0 if there is no overlapping word between $s$ and $S^L_{r_t}$, since the $\mathit{imp}(w)$ values of all words in $s$ equal zero; otherwise $s$ is more similar with sentences in $S^L_{r_t}$ when it contains many words with high $\mathit{imp}(w)$ values.

\subsubsection{Feature-based analysis}
\label{sec:ml-analysis}
Feature-based analysis trains a statistical model with the sentences in $D^L$ and uses this model to predict the probability that a sentence is related to $t$.
It includes four main steps: training set preparation, feature engineering, model training, and prediction.

In this work, the training set is comprised of the sentences in labeled documents $D^L$.
The positive instances for training are the related sentences $S^L_{r_t}$ for an information type $t$.
Since the majority of sentences in $D^L$ are unrelated, to avoid extreme imbalance, we perform under-sampling over the unrelated sentences $S^L_{u_t}$ to form the negative instances.
We sample a subset of $S^L_{u_t}$ which are similar to $S^L_{r_t}$ because these sentences are expected to be more difficult to distinguish from $S^L_{r_t}$.
The similarity is measured using group similarity, since it reflects the overall similarity between $S^L_{r_t}$ and a sentence in $S^L_{u_t}$.
We then use ML to learn the actual distinguishing criteria.
More specifically, we calculate the similarity between the group vector of $S^L_{r_t}$ and the context vector of each sentence in $S^L_{u_t}$;
we select the top-ranked unrelated sentences according to the size of $S^L_{r_t}$.
These sentences are textually similar with $S^L_{r_t}$.
We remark that, given the writing style adopted in financial documents,
sentences close to each other (e.g., in the same paragraph) usually discuss similar topics.
This means that within a same paragraph that could be both related and unrelated sentences.
For this reason, a sentence in $S^L_{u_t}$ is also considered as a negative instance if it is the previous or the next sentence of a related sentence.

\begin{table*}[tbp]
	\caption{Sentence Features for Feature-based Analysis}
	\centering
	\footnotesize
	\resizebox{\textwidth}{!}{
		\begin{filecontents*}{data/feature-info.csv}
ID& 			     Details
\multirow{3}{*}{F1--10}	&	\textbf{(N)} Word importance \textbf{(T)} Float \textbf{(D)} We rank words by their word importance (see Section~\ref{sec:ir-analysis}). We select the top-10 words as 10 features. The value of a feature is 0 if the sentence does not contain the corresponding word; otherwise, the feature value is the word importance. \textbf{(I)} These words are more important for an information type $t$.

\multirow{2}{*}{F11--20} &\textbf{(N)} Phrase importance \textbf{(T)} Float \textbf{(D)} We pair every two adjacent words in a sentence as a phrase. Similar to F1--10, we select the top-10 most important phrases as features. \textbf{(I)} These phrases are more important for $t$.

\multirow{2}{*}{F21}&\textbf{(N)} Length of a sentence \textbf{(T)} Integer \textbf{(D)} We count the number of words in a sentence. \textbf{(I)} Some information types are usually expressed with short sentences, e.g., ’Net Asset Value is calculated daily" (that explains the calculation frequency for issue price)}.

\multirow{2}{*}{F22}&\textbf{(N)} Ratio of `numbers' \textbf{(T)} Float \textbf{(D)} We count how many `numbers' in a sentence; the value is divided by the length of the sentence. \textbf{(I)} If most of the words in a sentence are numbers, the sentence is less likely to be related.

\multirow{2}{*}{F23}&\textbf{(N)} Number of `person name' \textbf{(T)} Integer \textbf{(D)} We count the number of `person names' in a sentence. \textbf{(I)} Some information types are associated with specific names, e.g., `corporate name'.

\multirow{2}{*}{F24}&\textbf{(N)} Number of `date' \textbf{(T)} Integer \textbf{(D)} We count the number of `date' in a sentence. \textbf{(I)} Some information types are associated with dates, e.g., `indication of date of establishment'.

\multirow{2}{*}{F25}&\textbf{(N)} Number of `web address' \textbf{(T)} Integer \textbf{(D)} We count the number of `web address' in a sentence. \textbf{(I)} Some information types may mention certain web addresses, e.g., `disclaimer on periodical reports' may mention the website to retrieve the reports.
\end{filecontents*}
\pgfplotstabletypeset[
	every head row/.style={
		before row=\toprule,
		after row=\midrule,
	},
	every nth row={1}{before row=\midrule},
	every last row/.style={
		after row=\bottomrule},
	columns/ID/.style ={column name={\textbf{ID}}}, 
	columns/Details/.style ={column name={\textbf{Name (N), Type (T), Description (D), and Intuition (I)}}, column type={p{.9\textwidth}}}, 	
	col sep=ampersand,
	string type,
	]{data/feature-info.csv}

 	}
	\label{tab:features}
\end{table*}

Regarding feature engineering, we constructed 25 features for model training. Table~\ref{tab:features} shows the name, type, and description of these features, and also their rationale.
F1--10 and F11--20 are 20 features related to the important words and phrases for an information type $t$.
F21--F25 calculate the length of a sentence and different types of named entities that could indicate the presence of an information type.
With these features, each sentence in the training set can be transformed into a feature vector.
The label of the feature vector is 1 or 0, representing whether a sentence is related to $t$ or not.

In this step, we construct features based on textual information of financial documents.
We do not use structural information
(e.g., section headings and font formatting)
for feature engineering because of two reasons.
First, structural information is not always parseable.
For readability reasons, in practice, a large number of financial documents are
made available in Portable Document Format (PDF).
A lot of structural information in such documents is missing~\cite{abreu2019findse},
since the PDF format organizes text blocks based on the graphical coordinates of characters.
Second, structural information in financial documents can be volatile~\cite{boella2016eunomos, sannier2017legal}.
Companies might use their unique templates to prepare financial documents;
document templates and structures can change significantly across different companies and versions.
Hence, to boost the generality of \approach,
we do not use structural information and only analyze textual information,
which is always available in financial documents.

We trained a random forest model for each information type with its corresponding feature vectors.
Random forest constructs a multitude of decision trees. Each decision tree is trained on a randomly selected subset of the training set.
We use random forest because it is known to better address over-fitting on small datasets~\cite{friedman2001elements}; decision trees are also able to automatically identify the most discriminative features for an information type (i.e., feature selection).

During prediction,
we transform candidate sentences into feature vectors and feed each of them into the trained model for the information type $t$, thus obtaining the  probability of each sentence to be related to $t$ (denoted as $s.\mathit{prob}$).

\subsubsection*{Application to the running example}
\figurename~\ref{fig:analysis-flow} illustrates the way we compute similarity or probability values in the fine-grained sentence analysis.
In \figurename~\ref{fig:analysis-flow}(A), \approach transforms related sentences in $S^L_{r_t}$ into a group vector.
This group vector is compared with the context vector of each candidate sentence $s_i$ by cosine similarity to obtain $s.\mathit{group}$ for $s_i$.
In \figurename~\ref{fig:analysis-flow}(B), \approach analyzes each related sentence in $S^L_{r_t}$ independently.
\approach computes the cosine similarity between the context vectors of the candidate sentence $s_i$
and each related sentence to get $s.\mathit{avg\_s}$ and $s.\mathit{max\_s}$.
In \figurename~\ref{fig:analysis-flow}(C), \approach gets the importance of each word $\mathit{imp}(w)$ in the related sentences,
which is used to compute $s.\mathit{word}$ of each candidate sentence based on the word importance similarity.
After the similarity-based analysis, in \figurename~\ref{fig:analysis-flow}(D),
\approach constructs feature vectors for positive and negative instances selected from related sentences and unrelated sentences, respectively.
These feature vectors are used to train a random forest model,
which can output $s.\mathit{ml}$ based on the feature vector of each candidate sentence $s_i$.
An example of the outputs of this step is shown in Table~\ref{tab:selection-example},
where we compute similarity or probability values for five candidate sentences.

\subsection{Sentence Selection}
\label{sec:sentence-sel}

\begin{table}[tbp]
	\caption{Example of sentence similarity}
	\centering
	\begin{filecontents*}{data/selection.csv}
S,      Group, Avg,  Max,  Word, ML,   Total
$s_1$,  0.80,  0.76, 0.91, 0.39, 0.67, 0.71
$s_2$,  0.60,  0.76, 0.82, 0.45, 0.76, 0.68
$s_{10}$, 0.73,  0.66, 0.80, 0.56, 0.70, 0.69
$s_{12}$, 0.82,  0.73, 0.88, 0.65, 0.66, 0.75
$s_{16}$, 0.75,  0.76, 0.81, 0.49, 0.67, 0.70
\end{filecontents*}
\pgfplotstabletypeset[
	every head row/.style={
		before row=\toprule,
		after row=\midrule,
	},
every last row/.style={
		after row=\bottomrule},
		columns/S/.style ={column name={ID}}, 
		columns/Group/.style ={column name={$s.\mathit{group}$}},
		columns/Avg/.style ={column name={$s.\mathit{avg}$}},
		columns/Max/.style ={column name={$s.\mathit{max}$}},
		columns/Word/.style ={column name={$s.\mathit{word}$}},
		columns/ML/.style ={column name={$s.\mathit{ml}$}},
		columns/Total/.style ={column name={$s.\mathit{score}$}
	}, 
	col sep=comma,
	string type,
]{data/selection.csv}

 	\label{tab:selection-example}
\end{table}

\begin{algorithm}[tb]
	\small
	\caption{Selector}
	\label{alg:selector}
	\KwIn{Candidate sentences $S^d_{c_t}$
		with $s.\mathit{group}$, $s.\mathit{avg\_s}$, $s.\mathit{max\_s}$, $s.\mathit{word}$, and $s.\mathit{ml}$ of each sentence \newline
		Average number of related sentences per document $n_r$\newline
		Related phrase list $\mathit{rlist}$ and unrelated phrase list $\mathit{ulist}$ \newline
Threshold $\theta$ for highly similar sentences
	}
	\KwOut{Sentences $S$ in $d$ related to $t$}
	${S^d_{c_t}}^\prime \gets $\emph{mergeSimilar}$(S^d_{c_t}, \theta)$\;\label{line:detect-dup}
	$S \gets$\emph{selectBySimilarity}$({S^d_{c_t}}^\prime, \theta)$\;\label{line:sel-by-sim}
	${S^d_{c_t}}^\prime\gets {S^d_{c_t}}^\prime \backslash S$\;
	\ForEach{$s \in {S^d_{c_t}}^\prime$}{\label{line:score-s}
		$s.\mathit{score}\gets (s.\mathit{group} + s.\mathit{avg\_s} + s.\mathit{max\_s} + s.\mathit{word}+s.\mathit{ml})/5$\;\label{line:score-score}
	}\label{line:score-e}
	$S^{rank}_\mathit{c_t}\gets$\emph{rankByScore}$({S^d_{c_t}}^\prime)$\;\label{line:select-by-score-s}
	$i = 0$\;
	\While{$\lvert S \rvert<n_r $ and $ i < \lvert S^{rank}_{c_t} \rvert$}{
		$S = S \cup {S^{rank}_{c_t}}[i++]$\;
	}\label{line:select-by-score-e}
	$S_g\gets$\emph{groupByDistance}$(S)$\;
	$S\gets \emptyset$\;
	\ForEach{$s_g \in S_g$}{\label{line:list-check-s}
		Boolean $rCheck\gets$\emph{hasWordsInList}$(s_g, rlist)$\;
		Boolean $uCheck\gets$ \emph{hasWordsInList}$(s_g, ulist)$\;
		\If{$rCheck$ and $!uCheck$}{
			$S = \{s_g\} \cup S$\;
		}
	}\label{line:list-check-e}
	return $S$\;
\end{algorithm}

For an information type $t$, \approach selects sentences from a document $d$ according to Algorithm~\ref{alg:selector}.
The basic idea is to select sentences that are either highly similar to $S^L_{r_t}$ regarding at least one aspect (i.e., the overall level, the individual sentence level, or the important word level) or ranked higher by the comprehensive score decided by both similarity- and feature-based analysis.
If we combine the above with key-phrase lists, we can refine the selection of sentences and accurately decide the final related sentence set.
The inputs include the candidate sentences $S^d_{c_t}$ and their
similarity and probability scores, the average number of related
sentences per labeled document $n_r$, a list of domain-specific related
phrases $\mathit{rlist}$ and unrelated phrases $\mathit{ulist}$, and
the auxiliary parameter $\theta$.
Notice that $\mathit{rlist}$ summarizes the phrases frequently used to express $t$; $\mathit{ulist}$ contains the phrases that are synonyms with related phrases but are seldomly used to express $t$.
Since domain experts usually use keyword search to help them find the possible location of related sentences, these lists can be manually constructed when deciding the criteria to annotate the training documents $D^L$.

Before sentence selection,
\approach detects duplicate sentences in $S^d_{c_t}$ (line~\ref{line:detect-dup}).
Two sentences are considered as duplicates if the similarity between their context vectors is larger than a threshold $\theta$.
Duplicate sentences are similar and usually express the same semantic meaning.
\approach will either select or exclude them together.

\approach first selects sentences by their similarity scores.
A sentence (and its duplicates) is selected if at least one of its similarity score ($s.\mathit{group}$, $s.\mathit{avg\_s}$, $s.\mathit{max\_s}$, or $s.\mathit{word}$) is greater than $\theta$,
because this sentence may express the same meaning as some related sentences in $D^L$.
As for unselected sentences, we calculate a sentence score for each sentence according to its similarity (from similarity-based analysis) and probability (from feature-based analysis) scores  (lines~\ref{line:score-s}--\ref{line:score-e}).
We rank sentences by their sentence scores and select the top-ranked sentences (and their duplicates) until the number of selected sentences reaches $n_r$ (lines~\ref{line:select-by-score-s}--\ref{line:select-by-score-e}).

Last, we group the selected sentences based on their position in the document, because information types are usually addressed by several continuous sentences.
We put any two sentences into a group if the distance between them is
less than three sentences, since these sentences usually share the same context.
For example, we put sentences $s_i$ and $s_{i+2}$ into a group as they have the same context sentence $s_{i+1}$.
For each group of sentences, we check whether they contain domain-specific phrases in the related list $\mathit{rlist}$ or unrelated list $\mathit{ulist}$.
We annotate a group of sentences as related if they feature phrases in $\mathit{rlist}$ but no phrase in $\mathit{ulist}$ (lines~\ref{line:list-check-s}--\ref{line:list-check-e}).

\subsubsection*{Application to the running example}
In this step, we assume to configure \approach to select four
sentences from the five candidate sentences listed in Table~\ref{tab:selection-example}.
First, \approach detects duplicate sentences.
In this example, no sentence pairs are duplicate.
Then, \approach selects sentences by their similarity scores.
In this example, we set the threshold $\theta$ to 0.9.
Therefore, $s_1$ is selected.
For the remaining four sentences, we rank them based on $s.\mathit{score}$,
and select the top-three sentences (i.e., $s_{12}$, $s_{16}$, and $s_{10}$).
Third, \approach groups the selected four sentences based on their position.
We get three groups, which are $s_1$, $s_{10}$--$s_{12}$, and $s_{16}$.
\approach decides the final sentences by comparing words in each group with $\mathit{rlist}$ and $\mathit{ulist}$.

\section{Evaluation}
\label{sec:evaluation}

In this section, we evaluate our approach
(\approach) for financial information type identification. First, we assess the
accuracy of \approach in identifying the sentences related to an information type. Then, we evaluate the factors that impact this accuracy, including different AI techniques (i.e., similarity-based analysis with IR and feature-based analysis with ML), the size of the training set, and the sentence selection strategy. Last, we analyze how \approach helps inspect financial documents.
More specifically, we answer the following research questions:
\begin{compactenum}[RQ1]
	\item \emph{Can \approach accurately identify the information types in financial documents?}
	\item \emph{How do different AI techniques affect the accuracy of \approach?}
	\item \emph{What is the impact of the size of labeled documents on the accuracy of \approach?}
	\item \emph{What is the impact of the sentence selection strategy on the accuracy of \approach?}
	\item \emph{How can \approach support the compliance analysis of financial documents?}
\end{compactenum}

\subsection{Dataset and Settings}
\label{sec:eval:dataset}

\begin{table}[tbp]
	\caption{Basic Statistics on the Dataset}
	\centering
	\begin{filecontents*}{data/dataset-info.csv}
Key,   Value
Num. of prospectuses , 100
Years of publishing, 2010--2021
Num. of pages (avg. / range) , 119 / 36--547
Num. of sentences (avg. / range), 2968 / 683--15458
\midrule, \multicolumn{1}{l}{T1: \hspace{0.5em} 3.6 / 1--12}
Num. of sentences, \multicolumn{1}{l}{T2: \hspace{0.5em} 5.6 / 1--29}
annotated, \multicolumn{1}{l}{T3: \hspace{0.03em} 17.8 / 11--54}
(avg. / range), \multicolumn{1}{l}{T4: \hspace{0.03em} 15.9 / 6--27}
, \multicolumn{1}{l}{T5: \hspace{0.03em} 36.4 / 4--97}
Num. of phrases in $\mathit{rlist}$, 151
Num. of phrases in $\mathit{ulist}$, 82
\end{filecontents*}
\pgfplotstabletypeset[
	every head row/.style={
		before row=\toprule,
		after row=\midrule,
	},
every last row/.style={
		after row=\bottomrule},
		columns/Key/.style ={column name={Item}}, 
		columns/Value/.style ={column name={Value}
	}, 
	col sep=comma,
	string type,
]{data/dataset-info.csv}

 	\label{tab:dataset}
\end{table}

We evaluated \approach with the content requirements for UCITS prospectuses~\cite{cssf2010the},
because UCITS is one of the most popular and representative investment regulatory frameworks, which has over \euro{10} trillion of assets under management across the world~\cite{ucits10tn}.

We randomly collected 100 approved UCITS prospectuses from the official website of a national regulator\footnote{CSSF approved prospectuses. \url{https://www.bourse.lu/home}} as our dataset.
The basic statistics on the dataset are shown in Table~\ref{tab:dataset}.
The dataset contains prospectuses published between 2010 and 2021, covering
all the calendar years
since the establishment of the ``UCITS Law 2010''.
The number of pages of these prospectuses varies significantly from 36 to 547,
with an average of 119.
In these prospectuses, the number of sentences ranges from 683 to \num[mode=text]{15458}.
On average each prospectus has \num[mode=text]{2968} sentences.

All the documents were annotated by three domain experts. Due to the
time required for annotating documents,
the domain experts selected five representative information types for
evaluation, including \emph{disclaimer on periodical reports} (T1),
\emph{calculation frequency for issue price} (T2), \emph{calculation
  method for issue price} (T3), \emph{liquidation conditions and
  procedure} (T4), and \emph{issue conditions and procedure} (T5).
They selected these information types by considering their importance, complexity,
and diversity. First, all these information types are content requirements,
which require to be present --- with explanatory sentences --- in every prospectus.
Second, manually identifying
these information types is time-consuming, as one has to identify relevant
sentences among, on average, around 3000 sentences.
Third, these information
types are diverse in terms of the average number of related sentences
(as shown in  Table~\ref{tab:dataset}, we have 3.6 for T1, 5.6 for T2, 17.8 for T3, 15.9 for T4, and 36.4 for
T5).
For example, the information type T1 has at most 12 related sentences per prospectus;
in contrast, the information type T5 can be related to up to 97 sentences.
These information types are also diverse in terms of the wording and writing styles.
For example, T1 (\emph{disclaimer on periodical reports}) is usually explained by only a few sentences with key phrases illustrating the places and the charge to obtain the reports,
while T5 (\emph{issue conditions and procedure}) can be discussed with many long sentences,
demonstrating different conditions and procedures to process the issue
(e.g., reject subscriptions, limit/restrict the issue).
Overall, this selection allowed us to
assess how \approach  identifies information types specified at
different levels of details (i.e., from a single sentence to several
pages of sentences).

\begin{table*}[tbp]
	\caption{Issues to be solved during the two Phases of the
            Annotation Process}
    \small
	\centering
		\begin{filecontents*}{data/annotation-info.csv}
Phase& 			     Type&	Meaning&	Solution
\multirow{2}{.1\textwidth}{Phase 1: Determining annotation criteria}	&	Vague criterion & The description of an initial criterion is vague & Improve the text of the criterion to avoid misunderstandings
& Missing criterion& A criterion cannot cover all related sentences or may lead to the inclusion of some unrelated sentences & Add new inclusion or exclusion criteria for each information type

\multirow{2}{.1\textwidth}{Phase 2: Annotating information types}	&	Missing annotation & A part of related sentences is not annotated & Cross-check the annotations to identify the missing parts
& Unrelated annotation& Some sentences are partially related& Discuss whether to add or delete these sentences case by case; more annotation criteria can be created to ensure agreement among the annotators
\end{filecontents*}
\pgfplotstabletypeset[
		every head row/.style={
		before row=\toprule,
		after row=\midrule,
	},
	every nth row={2}{before row=\midrule},
	every last row/.style={
	after row=\bottomrule},
	columns/Phase/.style ={column name={\textbf{Phase}}, column type={p{.1\textwidth}}}, 
	columns/Type/.style ={column name={\textbf{Issue Type}}, column type={p{.15\textwidth}}}, 	
	columns/Meaning/.style ={column name={\textbf{Meaning}}, column type={p{.2\textwidth}}}, 
	columns/Solution/.style ={column name={\textbf{Solution}}, column type={p{.4\textwidth}}}, 
	col sep=ampersand,
	string type,
	]{data/annotation-info.csv}

 	\label{tab:annotation}
\end{table*}

The annotation was conducted in two phases. First, the domain experts
selected 50 documents from the dataset. They perused these documents
to define the detailed criteria for annotation (i.e., what types of
sentences/phrases should be related/unrelated to an information
type).
In this phase, one of the domain experts first defined the
  initial annotation criteria.
For example, the criteria to annotate a sentence
as T1:\emph{disclaimer on periodical reports} are:
\begin{inparaenum}[(1)]
	\item ``the sentence  indicates where the periodical reports may be obtained
	(e.g., a website or the registered office of the management company)'';
	\item  ``the sentence  indicates that the reports can be obtained free of charge'';
	\item ``the sentence  describes the periodical reports (e.g., their frequency) and is in the same paragraph with one of the sentences satisfying  criterion (1) or (2)''.
\end{inparaenum}
The expert defined and consolidated the initial criteria over
  two weeks; this time frame includes the time to define the individual
  annotation criteria for each information type as well as the time to resolve
  interdependencies among criteria.
The initial criteria were then sent to the other two domain experts.
They had four 2-hour workshops in two weeks to further
refine the criteria. The main issues discussed during such
workshops were \emph{vague criteria} and  \emph{missing criteria}. As shown in the top row of
Table~\ref{tab:annotation}, vague criteria (i.e., criteria for which
the initial description was vague) were refined by improving the
their textual description; when the experts noticed that a
criterion could not cover all related sentences or could lead to the
inclusion of some
unrelated sentences for a given information type, they added new inclusion or exclusion criteria for that information type.
For example, for T1, they added the exclusion criterion ``we do not annotate sentences that describe the periodical reports but  are in different paragraphs from the sentences  satisfying criterion (1) or (2)'' to avoid false positive annotations.
Overall, the first phase of annotation took about one month.

During the second phase of the annotation, domain experts annotated all 100
documents with the selected information types based on the established
annotation criteria. Each person annotated a disjoint subset of
documents individually.
In this phase, we allocated three weeks for all the domain experts to complete their initial annotation.
They then examined each other's annotations.
As part of this step, five 2-hour workshops were conducted over five
weeks to discuss possible incorrect annotations and fix them when warranted.
The two main types of issue detected in this step, shown in the
bottom part of Table~\ref{tab:annotation}, were
\emph{missing annotation} and \emph{unrelated annotation}.
Missing annotations occurred because some information types (e.g., T5)
were associated with dozens of sentences spread across different pages;
domain experts could easily miss to annotate some of these related
sentences. Such issues were solved by cross-checking the annotations
to identify the missing parts.
An unrelated annotation indicated that a domain expert had wrongly
marked some sentences as related; this happened because some
paragraphs were only partially related to an information type.
Issues of this type were resolved by discussing, case by case, whether
to add or delete sentences, and by adding or removing annotation
criteria to ensure an agreement among the annotators.
For example, in the case of T3:\emph{calculation method for issue
  price}, although the documents include some sentences mentioning ``issue price'',
they do not provide the details for the calculation.
Domain experts finally considered such sentences as unrelated,
and added a new exclusion criterion for T3 ``we do not capture
sentences generically indicating how the sales or subscription price,
or the net asset value, are calculated, if they do not provide --- or
reference --- the details of the calculation''.
Overall, the second phase of annotation took in total about two months.

We remark that we did not consider a larger number of information
types due to the complexity of and the time required by the annotation
task.  On the one hand, domain experts have to identify the related
sentences from thousands of sentences in a prospectus.  On the other
hand, an information type can be related to multiple sentences, which
are usually distributed across different pages.  This means that domain experts
need to identify and double-check all these sentences.

The phrase lists $\mathit{rlist}$ and $\mathit{ulist}$  (see
\S~\ref{sec:sentence-sel}) used for sentence selection were manually
built based on the phrases listed in the annotation criteria. We extended the
phrases with synonyms occurring in the first 50 documents.
There are in total 151 phrases in the $\mathit{rlist}$ and 82 phrases in the $\mathit{ulist}$.
We set the parameter $\theta$ to 0.9; it was decided empirically by
evaluating the accuracy of \approach using a range of values between 0.1
and 1 with a step of 0.1 on the first group of 50 documents.

We performed the experiments with a
computer running macOS 11.1 with a 2.30 GHz Intel Core
i9 processor and 32GB memory.

\subsection{Accuracy of \approach (RQ1)}
\label{sec:rq-accuracy}

\begin{table*}[tb]
	\caption{Accuracy of Information Type Identification Algorithms}
	\centering
	\footnotesize
	\resizebox{\textwidth}{!}{
	\pgfplotstabletypeset[
	every head row/.style={
		before row={\toprule
			\multirow{2.8}{*}{ID}&\multicolumn{4}{c}{Precision}&\multicolumn{4}{c}{Recall}&\multicolumn{4}{c}{F$_1$-score}\\
			\cmidrule(l){2-5}
			\cmidrule(lr){6-9}
			\cmidrule(r){10-13}
		},
		after row=\midrule,
	},
	every last row/.style={
		after row=\bottomrule
	},
	every nth row={5}{before row=\midrule},
	columns/ID/.style ={column name=}, 
	columns/P-KW/.style ={column name=KW}, 
	columns/P-BERT/.style ={column name=BERT}, 
	columns/P-BERTKW/.style ={column name=BERT$_\mathit{KW}$}, 
	columns/P-FITI/.style={column name=\approach}, 
	columns/R-KW/.style ={column name=KW},
	columns/R-BERT/.style ={column name=BERT}, 
	columns/R-BERTKW/.style ={column name=BERT$_\mathit{KW}$}, 
	columns/R-FITI/.style={column name=\approach},	 
	columns/F-KW/.style ={column name=KW},
	columns/F-BERT/.style ={column name=BERT}, 
	columns/F-BERTKW/.style ={column name=BERT$_\mathit{KW}$}, 
	columns/F-FITI/.style ={column name=\approach},
	col sep=comma,
	string type,
]{data/rq-accuracy.csv}

 	}
	\label{tab:rq-accuray}
\end{table*}

To answer RQ1, we assessed the accuracy of \approach in identifying sentences related to different information types.

\subsubsection{Methodology}

We evaluated \approach using the annotated documents with $k$-fold cross-validation ($k=5$).
Since 50 annotated documents were used for phrase list construction and  parameter tuning, we kept them in the training set and only test \approach on the remaining 50 documents.
In each fold, we selected 10 documents from the remaining 50 documents as the test set; the training set included the other 90 documents.
Given an information type $t$ and a test set,
we compared the sentences selected by \approach with the ground
truth annotated by the domain experts.
We measured the accuracy of \approach with
\emph{precision}, \emph{recall}, and \emph{F$_1$-score} ($F_1$).
They are defined as
$\mathit{Precision}=\frac{|\mathit{TP}|}{|\mathit{TP}|+|\mathit{FP}|}$
and
$\mathit{Recall}=\frac{|\mathit{TP}|}{|\mathit{TP}|+|\mathit{FN}|}$,
where true positives (TP) and false positives (FP) refer to sentences selected by \approach which are related or not to  $t$, respectively.
False negatives (FN) refer to cases where \approach misses a sentence related to  $t$.
\emph{F$_1$-score} is
defined as
$F_1=\frac{2\times \mathit{Precision} \times \mathit{Recall}  }{\mathit{Precision}+\mathit{Recall}}$.
These evaluation metrics work at the sentence level.
For example, when there are four consecutive sentences annotated by the domain experts in the ground truth, and \approach correctly identifies three of them,
we have $\mathit{TP}=3$,  $\mathit{FP}=1$, and $\mathit{Precision}= 0.75$.

We compared \approach with three baselines:
 a keyword-search strategy (denoted as KW),
the Bidirectional Encoder Representation from Transformers (BERT)
language model,
and a combination of BERT and KW (denoted as BERT$_\mathit{KW}$).

 KW is a common way for regulator agents to locate sentences.
	Since we have constructed two phrase lists containing related phrases (i.e., $\mathit{rlist}$) and unrelated phrases (i.e., $\mathit{ulist}$), KW directly analyze the candidate sentences to select those that feature phrases in $\mathit{rlist}$ and no phrase in $\mathit{ulist}$ as related to an information type.

       BERT is a transformer-based language model for NLP~\cite{devlin2018bert,rogers2020primer}.
	We take BERT as a baseline
	because it is a typical deep neural network that has been widely used for requirement mining in requirement engineering~\cite{wang2022detecting}
	The input of BERT is a sentence for classification and its context sentences
	(i.e., its previous sentence and the next sentence, as explained in Section~\ref{sec:candidates}),
	which is the same information used by \approach.
	In the training phase, we fine-tuned a pre-trained BERT for each information type
	with the training set constructed in Section~\ref{sec:ml-analysis} for feature-based analysis of \approach.
	In the testing phase, BERT predicted the relatedness of each candidate sentence to an information type.

       BERT$_\mathit{KW}$ is a combination of BERT and KW.
	For each sentence predicted as related by BERT,
	we improved the prediction based on key-phrase lists.
	Among the sentences predicted by BERT as related, we further retained the sentences that feature phrases in $\mathit{rlist}$ but no phrase in $\mathit{ulist}$ as the final set of sentences related to an information type.

We implemented KW from scratch.
BERT was implemented with the open source Python library for transformers\footnote{Hugging Face Transformers \url{https://github.com/huggingface/transformers}}.
BERT was pretrained with the \texttt{bert-base-uncased} dataset.
Hyper-parameters were fine-tuned using early stopping and the Adam optimizer based on the cross entropy loss~\cite{chalkidis2021regulatory}.
In each iteration of fine-tuning, we set the batch size to 64.
Fine-tuning stopped when the loss did not change for 10 iterations.

\subsubsection{Results}
\label{sec:rq1-results}

Table~\ref{tab:rq-accuray} shows the accuracy of the different information type identification algorithms.
\approach identified a set of sentences related to each information
type with a precision value ranging from \numrange{\FITIPLow}{\FITIPHigh}.
The differences in precision between these information types are less than 10\%.
The precision value for T4 is the highest (\FITIPHigh).
We found that prospectuses tend to use similar sentences to explain the liquidation conditions and procedure (T4);
therefore, all the algorithms got a relatively high precision score on this information type.
The precision values of T2 and T3 are relatively low, but still near
80\% for \approach.
The recall value of \approach ranges from \numrange{\FITIRLow}{\FITIRHigh}, with an average recall value of {\FITIRAvg}.
The result means that \approach could find, on average, more than 60\% of sentences for an information type.
Among these information types, the recall values of T2 and T5 are low.
For T2, the low recall value is caused by the small number of sentences related to T2.
When a related sentence is not identified by \approach,
recall could change dramatically.
For T5, there are on average 36.4 sentences related.
\approach may not easily identify all of these sentences,
leading to low recall.

KW performed poorly compared to \approach, identifying an average of
{\KWRAvg} related sentences with an average precision value of
{\KWPAvg}.  \approach outperformed KW by 0.277 ({\FITIFAvg} vs
{\KWFAvg}) in terms of F$_1$-score.  Although keyword search is a
common activity for regulator agents, many false positives can be
returned due to the case that sentences related to different information types share a common vocabulary. Moreover, since the writing styles and the number of related sentences can differ across financial documents from different investment companies, regulator agents may not enumerate all keywords
for every related sentence, leading to a low recall value.  In contrast, \approach could leverage these (possibly incomplete) keyword
lists to improve its accuracy in identifying information types.

As for the transformer language model BERT,
its average recall value is {\BERTRAvg},
which is similar to the one obtained by \approach.
BERT has a higher recall than \approach for information types T2--T5.
However, the precision value of BERT is much lower than that of \approach (i.e., {\BERTPAvg} vs {\FITIPAvg}),
since BERT wrongly predicts many sentences as related.
When integrating the key-phrase lists into BERT,
BERT$_\mathit{KW}$ can identify information types more accurately.
The average precision of BERT$_\mathit{KW}$ improves from {\BERTPAvg}
(the value achieved by BERT) to {\BERTKWPAvg}.
However, as a trade-off, the average recall of BERT$_\mathit{KW}$ drops dramatically to {\BERTKWRAvg}.
As a result, \approach outperforms both BERT and BERT$_\mathit{KW}$ in terms of F$_1$-score.
The results obtained by BERT can be explained as follows.
In this task, we need to account for the limited
size of domain-specific datasets, bounded by the high cost of
annotations, which must be performed by domain experts.
Numerous neural network weights in a deep neural network like BERT may not be well fine-tuned in this context.
In addition, as discussed in Section~\ref{sec:background},
financial documents can use similar sentences to explain different types of required information.
Hence, many sentences are classified by BERT as false positives, leading
to a low precision value.

We conducted the Wilcoxon test on the prediction outputs of the
  different algorithms in terms of the F$_1$-score obtained for each
  information type in the testing documents; we chose this test
since it is non-parametric and does not require any distributional assumption~\cite{demvsar2006statistical}.
The results confirm that the differences in the prediction between \approach and
the baselines are statistically significant (p-value $<$ 0.05).

The current results of \approach can be interpreted as follows.
With an average
precision above 80\%, \approach can help users efficiently
locate the correct position of 64.6\% related sentences.
Compared to analyzing the whole document, users can use \approach to reduce the effort
in manually finding sentences related to an information type.
The current results are also important for \approach to be used in the context
of (financial) document compliance checking.
In compliance checking the existence of an information type can be
established if  \approach can identify at least one sentence for this information type.
With a high precision value, it means that when \approach finds some sentences,
they are usually the actual related sentences for an information type.
Therefore, \approach can correctly decide the existence of this information type.
The high precision value is also important to detect missing information types.
\approach can confidently (i.e., with high precision) find related sentences for an information type.
These sentences are frequently used to explain an information type.
However, given a financial document,
if \approach cannot find any of such sentences,
it is more likely that the financial document misses this information type.

\paragraph{Performance analysis}\label{sec:rq1:case}
We have further analyzed the predictions made by \approach,
to identify the cases in which it performs well as well as those with subpar performance. We have identified four cases:

\textbf{Case 1 (Positive)}: \emph{\approach can correctly
    identify information types when they are explained with similar sentences in different prospectuses.}
We found that for some information types,
prospectus writers tend to use similar structures and sentences to explain them.
A typical example is T4: \emph{liquidation conditions and procedure},
where \approach and all baselines got the highest F$_1$-score on the dataset.
Since \approach conducts multi-granularity similarity analysis (i.e., group similarity, sentence similarity, and word importance similarity) on the documents,
the similarity across different documents can be correctly identified by \approach.

\textbf{Case 2 (Positive)}: \emph{\approach performs well when
    the sentences related to information types contain certain named entities or keywords.}
\approach uses feature-based analysis to capture important words, phrases, and named entities (i.e., number, person name, date, and web address).
When the sentences of an information type are associated with such keywords and named entities,
\approach could distinguish these sentences from those related to other information types.
For instance, information type T1:\emph{disclaimer on periodical reports} indicates where the periodical reports may be obtained.
It is usually associated with (web) addresses for obtaining the periodical reports.
The feature-based analysis of \approach can correctly identify these
sentences,  increasing prediction accuracy.

\textbf{Case 3 (Negative)}: \emph{\approach may select sentences
    that are only partially related to an information type.}
  As discussed in Section~\ref{sec:eval:dataset},
some paragraphs may be partially related to an information type,
which means that they mention an information type but do not provide
all the corresponding details.
These partially related sentences (e.g., T3:\emph{calculation method
  for issue price}) are difficult to label even for domain experts.
\approach may also include these partially related sentences in the prediction results,
leading to false positives.
A structure-based analysis of the documents, to be conducted as part
of future work, could reduce the false positives.
Although a concept (e.g., issue price) can be discussed in different sections,
some sections (e.g., the background) will only discuss an overview of the concept without going into the details.
By analyzing the structure of prospectuses,
some false positives could be filtered.

\textbf{Case 4 (Negative)}:
\emph{It is difficult for \approach to identify the sentences related to similar information types.}
Some information types are similar,
such as T2:\emph{calculation frequency for issue price}
and T3:\emph{calculation method for issue price}.
They are both related to ``issue price'', though the focus of each of
them is on different topics (i.e., calculation frequency and calculation method).
Since these two information types are related and usually explained together,
\approach may not correctly distinguish between them,
leading to low accuracy compared to other information types (as shown in Table~\ref{tab:rq-accuray}).
To increase the accuracy,
as part of future work,
topic models can be used to analyze the  different topics discussed in similar information types.

To conclude, \emph{the answer to RQ1 is that \approach identifies an average of
  64.6\% of relevant sentences for an information type, with an
  average precision value of {\FITIPAvg}, significantly outperforming
  the baselines based on keywords and language models.}

\subsection{Impact of Different Components (RQ2)}
\label{sec:rq-components}

\approach selects related sentences based on both
similarity- and feature-based analyses with IR and ML techniques (section~\ref{sec:sentence-sel}). To answer RQ2, we assessed the impact of these two types of analysis on the
accuracy of \approach.

\subsubsection{Methodology}

We implemented two variants of
\approach (called $\text{\approach}_\mathit{IR}$ and $\text{\approach}_\mathit{ML}$) to assess the possible impact of each technique.
During sentence selection, $\text{\approach}_\mathit{IR}$ only selects
sentences based on the similarity values calculated in the similarity-based
analysis, while $\text{\approach}_\mathit{ML}$ performs sentence
selection only relying on the probability calculated in the feature-based analysis.
To implement $\text{\approach}_\mathit{IR}$, we calculated the score of a sentence by averaging the four similarity values (i.e., $s.\mathit{group}$, $s.\mathit{avg\_s}$, $s.\mathit{max\_s}$, $s.\mathit{word}$) at line~\ref{line:score-score} in Algorithm~\ref{alg:selector}.
To implement $\text{\approach}_\mathit{ML}$, we disabled the function \emph{selectBySimilarity} at line~\ref{line:sel-by-sim} and assigned the score of a sentence as the probability value calculated from the feature-based analysis ($s.\mathit{ml}$).
We ran the standard version of \approach (i.e., the one presented in
section~\ref{sec:algorithm}) and the additional variants using the same settings
as in RQ1.

\subsubsection{Results}

\begin{figure*}[tbp]\centering
	
\subfloat[Precision of \approach and its variants]{
	\label{fig:rq-component-p}{
		\begin{tikzpicture}
		[scale=0.46]
		\begin{axis} [ybar = .05cm,
		bar width = 6pt,
		xmin = 0,
		xmax = 6,
		ymax = 1,
		enlarge x limits = {value = .25},
		xticklabels={T1, T2, T3, T4, T5},
		xtick={1,...,5},
		ylabel={Precision},
		ylabel near ticks,
                axis y discontinuity=parallel,
		legend pos=north west,
		]

		\addplot table [x=ID, y=FITI-IR-P, col sep=comma] {data/rq-component.csv};\addlegendentry{$\text{\approach}_\mathit{IR}$};
		\addplot table [x=ID, y=FITI-ML-P, col sep=comma] {data/rq-component.csv};\addlegendentry{$\text{\approach}_\mathit{ML}$};
		\addplot table [x=ID, y=FITI-P, col sep=comma] {data/rq-component.csv};
		\addlegendentry{\approach};

		\end{axis}
		\end{tikzpicture}
	}
}
\hfill
\subfloat[Recall of \approach and its variants]{
	\label{fig:rq-component-r}{
		\begin{tikzpicture}
		[scale=0.46]
		\begin{axis} [ybar = .05cm,
		bar width = 6pt,
		xmin = 0,
		xmax = 6,
		ymax = 1,
		enlarge x limits = {value = .25},
		xticklabels={T1, T2, T3, T4, T5},
		xtick={1,...,5},
		ylabel={Recall},
		ylabel near ticks,
                axis y discontinuity=parallel,
		legend pos=north west,
		]

		\addplot table [x=ID, y=FITI-IR-R, col sep=comma] {data/rq-component.csv};\addlegendentry{$\text{\approach}_\mathit{IR}$};
		\addplot table [x=ID, y=FITI-ML-R, col sep=comma] {data/rq-component.csv};\addlegendentry{$\text{\approach}_\mathit{ML}$};
		\addplot table [x=ID, y=FITI-R, col sep=comma] {data/rq-component.csv};
		\addlegendentry{\approach};

		\end{axis}
		\end{tikzpicture}
	}
}
\hfill
\subfloat[F$_1$-score of \approach and its variants]{
	\label{fig:rq-component-f}{
		\begin{tikzpicture}
		[scale=0.46]
		\begin{axis} [ybar = .05cm,
		bar width = 6pt,
		xmin = 0,
		xmax = 6,
		ymax = 1,
		enlarge x limits = {value = .25},
		xticklabels={T1, T2, T3, T4, T5},
		xtick={1,...,5},
		ylabel={F$_1$-score},
		ylabel near ticks,
                axis y discontinuity=parallel,
		legend pos=north west,
		]

		\addplot table [x=ID, y=FITI-IR-F, col sep=comma] {data/rq-component.csv};\addlegendentry{$\text{\approach}_\mathit{IR}$};
		\addplot table [x=ID, y=FITI-ML-F, col sep=comma] {data/rq-component.csv};\addlegendentry{$\text{\approach}_\mathit{ML}$};
		\addplot table [x=ID, y=FITI-F, col sep=comma] {data/rq-component.csv};
		\addlegendentry{\approach};

		\end{axis}
		\end{tikzpicture}
	}
      }

 	\caption{Impact of similarity-based and feature-based analyses for \approach}\label{fig:rq-component}\end{figure*}

As presented in Figure~\ref{fig:rq-component}, similarity- and feature-based analysis show different abilities in analyzing information types.

Feature-based analysis ($\text{\approach}_\mathit{ML}$) tends to assign a high probability value to a small fraction of related sentences.
Hence, $\text{\approach}_\mathit{ML}$ achieves higher precision values, for the majority of information types (i.e., T2 to T5), than $\text{\approach}_\mathit{IR}$.
The precision values of T2, T3, and T4 are also higher than those of \approach (in Figure~\ref{fig:rq-component-p}).
However, the recall value of $\text{\approach}_\mathit{ML}$ is lower than both $\text{\approach}_\mathit{IR}$ and \approach (in Figure~\ref{fig:rq-component-r}).

The above results can be explained as follows.
$\text{\approach}_\mathit{ML}$ analyzes sentences based on features related to important words/phrases, length of sentences, and named entities (as shown in Table~\ref{tab:features}).
Since these features capture the characteristics of information types,
$\text{\approach}_\mathit{ML}$ identifies a subset of related sentences
that best reflect these features with high precision.
For example, $\text{\approach}_\mathit{ML}$ can identify sentences containing the information of calculation date and important words for information type T2:\emph{calculation frequency for issue price}.
An exception is T1, for which the precision value is low.
We conjecture the reason is the small number of positive instances (i.e., 3.6 related sentences per document on average) for training the ML model.

As to similarity-based analysis ($\text{\approach}_\mathit{IR}$), it tends to retrieve more related sentences than $\text{\approach}_\mathit{ML}$, leading to a high recall, though some are false positives.
$\text{\approach}_\mathit{IR}$ is an instance-based algorithm
that retrieves many sentences similar to ground truth sentences.
However, as discussed in Section~\ref{sec:background},
financial documents may use similar sentences to explain different types of required information.
Many unrelated similar sentences could also be included,
leading to a high recall value but a low precision value.

By integrating the two components, the accuracy in identifying information types is further improved: \approach achieves a higher F$_1$-score than both $\text{\approach}_\mathit{IR}$ and $\text{\approach}_\mathit{ML}$ with p-value $<$ 0.05 regarding the five information types (in Figure~\ref{fig:rq-component-f}).
When integrating the two components,
more related sentences identified by either IR or ML are included, as well as some unrelated sentences
(e.g., similar sentences that explain other information types).
Hence, the final recall value is improved,
but the precision value is sometimes lower than that obtained by the other variants.
These results show that the integration of the components is necessary.
After integration, \approach achieves a substantial improvement in terms of recall compared with $\text{\approach}_\mathit{IR}$ and $\text{\approach}_\mathit{ML}$,
with an average precision value that is only less than 2\% lower than that of $\text{\approach}_\mathit{ML}$.

\emph{The answer to RQ2 is that both similarity-based and feature-based analyses contribute to improving the accuracy of
\approach, complementing each other.}

\subsection{Impact of the Size of the Training Set (RQ3)}
\label{sec:rq:size-impact}

\approach conducts information type identification relying on the  documents in the training set annotated by domain experts.
This RQ assesses the impact of the size of the training set on the accuracy of \approach. It is an important question as access to such annotated documents is, in practice, inherently limited.

\subsubsection{Methodology}\label{sec:rq3:method}
We evaluated the accuracy of \approach by varying the size of the
training set from \numrange{10}{90} documents in steps of $10$.
For each training set size, we built the
training set incrementally: each time we randomly selected 10
documents and added them to the training set. In other words, for
each training set size, the training set is a
superset of the one used for the previous value. For instance, the training
set for size 40 was obtained by randomly selecting 10 documents
and adding them to the training set for size 30.
We trained \approach on each sampled training set and used the trained model to identify sentences in the test set for different information types.
We measured the accuracy of \approach for different training set
sizes, as when addressing RQ1.
We used 5-fold cross validation, so we repeated this process five
times for each training set size.

\subsubsection{Results}\label{sec:rq3:results}

\begin{figure*}[tb]\centering
	
\subfloat[Precision of \approach]{
	\label{fig:rq-train-size-p}
	\begin{tikzpicture}[scale=0.46]
		\begin{axis}[
			xlabel={Number of training documents},
			ylabel={Precision},
			ylabel near ticks,
			xmin=0, xmax=100,
			legend style={
				at={(0.5,-0.2)},
				anchor=north,
				legend columns=-1
			},
			ymajorgrids=true,
                        axis y discontinuity=parallel,
			grid style=dashed,
			cycle list={
				solid, every mark/.append style={solid, fill=red}, mark=*\\dotted, every mark/.append style={solid, fill=green}, mark=square*\\densely dotted, every mark/.append style={solid, fill=blue}, mark=otimes*\\loosely dotted, every mark/.append style={solid, fill=cyan}, mark=triangle*\\dashed, every mark/.append style={solid, fill=yellow},mark=diamond*\\loosely dashed, every mark/.append style={solid, fill=olive},mark=*\\densely dashed, every mark/.append style={solid, fill=orange},mark=square*\\dashdotted, every mark/.append style={solid, fill=gray},mark=otimes*\\dasdotdotted, every mark/.append style={solid, fill=brown},mark=star\\densely dashdotted,every mark/.append style={solid, fill=teal},mark=diamond*\\},
			]

			\addplot table [x=ID, y=T1, col sep=comma] {data/rq-train-size-p.csv};\addlegendentry{T1};
			\addplot table [x=ID, y=T2, col sep=comma] {data/rq-train-size-p.csv};\addlegendentry{T2};
			\addplot table [x=ID, y=T3, col sep=comma] {data/rq-train-size-p.csv};
			\addlegendentry{T3};
			\addplot table [x=ID, y=T4, col sep=comma] {data/rq-train-size-p.csv};
			\addlegendentry{T4};
			\addplot table [x=ID, y=T5, col sep=comma] {data/rq-train-size-p.csv};
			\addlegendentry{T5};

		\end{axis}
	\end{tikzpicture}
}
\hfill
\subfloat[Recall of \approach]{
	\label{fig:rq-train-size-r}
	\begin{tikzpicture}[scale=0.46]
		\begin{axis}[
xlabel={Number of training documents},
		ylabel={Recall},
		ylabel near ticks,
		xmin=0, xmax=100,
		legend style={
			at={(0.5,-0.2)},
			anchor=north,
			legend columns=-1
		},
		ymajorgrids=true,
                axis y discontinuity=parallel,
		grid style=dashed,
		cycle list={
			solid, every mark/.append style={solid, fill=red}, mark=*\\dotted, every mark/.append style={solid, fill=green}, mark=square*\\densely dotted, every mark/.append style={solid, fill=blue}, mark=otimes*\\loosely dotted, every mark/.append style={solid, fill=cyan}, mark=triangle*\\dashed, every mark/.append style={solid, fill=yellow},mark=diamond*\\},
		]

		\addplot table [x=ID, y=T1, col sep=comma] {data/rq-train-size-r.csv};\addlegendentry{T1};
		\addplot table [x=ID, y=T2, col sep=comma] {data/rq-train-size-r.csv};\addlegendentry{T2};
		\addplot table [x=ID, y=T3, col sep=comma] {data/rq-train-size-r.csv};
		\addlegendentry{T3};
		\addplot table [x=ID, y=T4, col sep=comma] {data/rq-train-size-r.csv};
		\addlegendentry{T4};
		\addplot table [x=ID, y=T5, col sep=comma] {data/rq-train-size-r.csv};
		\addlegendentry{T5};

		\end{axis}
	\end{tikzpicture}
}
\hfill
\subfloat[F$_1$-score of \approach]{
	\label{fig:rq-train-size-f}
	\begin{tikzpicture}[scale=0.46]
		\begin{axis}[
		xlabel={Number of training documents},
		ylabel={F$_1$-score},
		ylabel near ticks,
		xmin=0, xmax=100,
		legend style={
			at={(0.5,-0.2)},
			anchor=north,
			legend columns=-1
		},
		ymajorgrids=true,
                axis y discontinuity=parallel,
		grid style=dashed,
		cycle list={
			solid, every mark/.append style={solid, fill=red}, mark=*\\dotted, every mark/.append style={solid, fill=green}, mark=square*\\densely dotted, every mark/.append style={solid, fill=blue}, mark=otimes*\\loosely dotted, every mark/.append style={solid, fill=cyan}, mark=triangle*\\dashed, every mark/.append style={solid, fill=yellow},mark=diamond*\\},
		]

		\addplot table [x=ID, y=T1, col sep=comma] {data/rq-train-size-f.csv};\addlegendentry{T1};
		\addplot table [x=ID, y=T2, col sep=comma] {data/rq-train-size-f.csv};\addlegendentry{T2};
		\addplot table [x=ID, y=T3, col sep=comma] {data/rq-train-size-f.csv};
		\addlegendentry{T3};
		\addplot table [x=ID, y=T4, col sep=comma] {data/rq-train-size-f.csv};
		\addlegendentry{T4};
		\addplot table [x=ID, y=T5, col sep=comma] {data/rq-train-size-f.csv};
		\addlegendentry{T5};

		\end{axis}
	\end{tikzpicture}
}

 	\caption{Impact of the size of the training set for \approach}\label{fig:rq-train-size}\end{figure*}

As shown in Figure~\ref{fig:rq-train-size},  precision
largely increases (i.e., in the case of T1) or becomes relatively stable
within a certain range (i.e., in the case of T2, T3, T4, and T5) as the size of the training set increases.
We found that with \numrange{10}{90} documents in the training set,
the precision for information types T1 and T2 fluctuates more;
as shown in Table~\ref{tab:dataset}, there are only \numrange{3.6}{5.6} sentences on average related to these information types.
When a related or an unrelated sentence is added to the prediction results,
the precision can thus change a lot.
For T3--T5, the fluctuation on precision is small.
The changes in precision (e.g., at the 70--80 documents threshold) are mainly caused by two reasons.
First, we randomly added 10 more documents to the training set each time;
the selection of these new documents affects  precision.
Second, when \approach selects more sentences to increase recall, as a complementary evaluation metric, precision on the 10 documents in the testing set  could be affected.

With more than 40 annotated documents, precision is stable, varying within a 10\% range for all information types.
Further, we observe that the precision value is still higher than 70\% for most information types (i.e., in the case of T2, T3, T4, and T5) when there are only 10 annotated documents.
This is because \approach is able to identify a small fraction of sentences that are highly similar to the annotated related sentences in the training set.
However, as shown in Figure~\ref{fig:rq-train-size-r}, the recall value of \approach is low; for information types T2, T3, and T5, recall is around 40\%.
As more documents are annotated, recall largely increases, because \approach may learn more about different expressions or wordings to explain an information type.
F$_1$-score also confirms the positive impact of increasing the size of the training set on \approach.

\emph{The answer to RQ3 is that the accuracy of
\approach improves as the size of the training set increases. \approach requires a minimum size of 40 annotated documents to achieve high and stable precision.}

\subsection{Impact of the Sentence Selection Strategy (RQ4)}
\label{sec:eval-selection}

\begin{table*}[tb]
	\caption{Accuracy of Different Sentence Selection Strategies}
	\centering
	\scriptsize
	\resizebox{\textwidth}{!}{
	\pgfplotstabletypeset[
	every head row/.style={
		before row={\toprule
			\multirow{2.5}{*}{ID}&\multicolumn{4}{c}{Precision}&\multicolumn{4}{c}{Recall}&\multicolumn{4}{c}{F$_1$-score}\\
			\cmidrule(l){2-5}
			\cmidrule(lr){6-9}
			\cmidrule(r){10-13}
		},
		after row=\midrule,
	},
	every last row/.style={
		after row=\bottomrule
	},
	every nth row={5}{before row=\midrule},
	columns/ID/.style ={column name=}, 
	columns/P-SC/.style ={column name=\approach{$_{\mathit{score}}$}}, 
	columns/P-TN/.style={column name=\approach{$_{\mathit{topN}}$}}, 
	columns/P-AVG/.style ={column name=\approach{$_{\mathit{avgN}}$}}, 
	columns/P-FITI/.style={column name=\approach}, 
	columns/R-SC/.style ={column name=\approach{$_{\mathit{score}}$}}, 
	columns/R-TN/.style={column name=\approach{$_{\mathit{topN}}$}}, 
	columns/R-AVG/.style ={column name=\approach{$_{\mathit{avgN}}$}}, 
	columns/R-FITI/.style={column name=\approach}, 
	columns/F-SC/.style ={column name=\approach{$_{\mathit{score}}$}}, 
	columns/F-TN/.style={column name=\approach{$_{\mathit{topN}}$}}, 
	columns/F-AVG/.style ={column name=\approach{$_{\mathit{avgN}}$}}, 
	columns/F-FITI/.style={column name=\approach}, 
	col sep=comma,
	string type,
]{data/rq-selection.csv}

 	}
	\label{tab:rq-selection}
\end{table*}

\approach selects sentences related to an information type based on
a list of domain-specific related phrases and a list of unrelated
phrases (\emph{rlist} and \emph{ulist} in
Algorithm~\ref{alg:financial-doc-checking}, respectively).
We expect these two lists to be manually built by domain experts
when defining the criteria for annotating the training set of
financial documents.
However, though a one-off task, it could cause significant burden to the
domain experts, in this RQ
we analyze alternative sentence selection strategies
to assess their impact on the accuracy of \approach.

\subsubsection{Methodology}

We compared the sentence selection strategy in \approach (which ranks
sentences based on their sentence scores
computed by the IR and ML techniques)
with three alternative strategies that are commonly used for selecting items from a ranked list:
\begin{itemize}
	\item \approach{$_{\mathit{score}}$} selects only sentences in
          a ranked list whose scores are above a user-defined threshold $\theta_\mathit{score}$;
	\item \approach{$_{\mathit{topN}}$}  selects the top-$N$ sentences in a ranked list, where $N$ is an input parameter;
	\item \approach{$_{\mathit{avgN}}$} sets a different $N$ for
          each information type, which is automatically determined
          based on the average number of sentences related to this information type in all financial documents in the training set; \approach{$_{\mathit{avgN}}$} uses this number to select the top-ranked sentences as related.
\end{itemize}

We replaced the sentence selection strategy in \approach with the
three above alternative strategies,
and compared the accuracy of such \approach variant for information type identification.
Since \approach{$_{\mathit{score}}$} and \approach{$_{\mathit{topN}}$} have additional input parameters
(i.e., $\theta_\mathit{score}$ and $N$, respectively),
we decided their values by conducting a preliminary evaluation, aimed
at identifying the parameter value of \approach{$_{\mathit{score}}$}
and \approach{$_{\mathit{topN}}$} leading to the highest
\emph{F$_1$-score} for the majority of information types.
We varied $\theta_\mathit{score}$ from 0 to 1 with a step of 0.05: the highest
\emph{F$_1$-score} was obtained when $\theta_\mathit{score}$ was set to 0.4.
We varied $N$ from 1 to 5 with a step of 1, obtaining the highest
\emph{F$_1$-score} when $N$ was set to 3.
We used these values ($\theta_\mathit{score}=0.4$ and $N=3$) in our experiments.

\subsubsection{Results}

Table~\ref{tab:rq-selection} shows the accuracy results (in terms of
precision, recall, and F$_1$-score) when using the
different sentence selection strategies;
the value in bold is the best result achieved for a given
evaluation metric.

From the table, we can see that the original sentence selection
strategy used by \approach (the one based on phrase lists) leads to
the best precision value for four out of five information types (and
yields the second-best result for the fifth information types, T4),
with an average precision value of {\FITIPAvg},
outperforming the three alternative strategies with a difference
ranging from 0.123 ({\TopNPAvg} vs {\FITIPAvg}) to 0.214 ({\AvgNPAvg}
vs {\FITIPAvg}).  The reason for such a high precision value is
that the phrase lists provide extra domain knowledge on related and
unrelated sentence patterns, which are important for \approach to
accurately assess how related (to a certain topic) a list of sentences
is.

Regarding recall, although the average recall value of \approach is
higher than those obtained when using the three alternative
strategies, \approach achieves the highest recall value only for one
information type (T4).  The reason is that, when leaving sentences
out based on the unrelated phrase list \emph{ulist}, a few related sentences
could be incorrectly dismissed.  The difference between the recall value of
\approach and the best recall value is less than 3\% for T1, T3, and
T5, and goes up to 10\% for T2.  As discussed in
Section~\ref{sec:rq-accuracy}, in the context of (financial) document
compliance checking, it is desirable to use identification algorithms
that yield a high precision value. Hence, a slight decrease in
recall is acceptable, considering the high precision value
achieved by \approach with the sentence selection strategy based on
phrase lists.  This is also reflected when looking at the overall
accuracy (i.e., the F$_1$-score), for which the original sentence
selection strategy outperforms all alternative
strategies with p-value $<$ 0.05.

\emph{The answer to RQ4 is that the sentence selection strategy used
  by \approach outperforms other alternative
  strategies. Such a strategy, based on  phrase lists, improves the precision of sentence selection.}

\subsection{\approach for Financial Document Compliance Checking (RQ5)}
\label{sec:appr-financ-docum}

To answer RQ5, we simulate the scenario of compliance checking for financial content requirements and analyze the accuracy of \approach in the context of such scenario.

\subsubsection{Methodology}
In our experiments, we assessed the accuracy of \approach in
identifying information types with a set of approved prospectuses,
which satisfy all content requirements.  However, in actual compliance checking cases, regulator agents usually inspect
(unapproved) prospectuses that do not fulfill some content requirement
(i.e., they lack some specific information).  We simulated this
scenario to understand how \approach can help regulator agents check
financial documents for compliance.

Since unapproved prospectuses are usually unavailable due to
confidentiality reasons, we simulated such prospectuses by removing
sentences related to an information type.  Specifically, we first
sampled 10 documents from the 50 testing documents as unapproved
prospectuses, to simulate the case where a small subset of documents is incomplete.  Second, we removed sentences
related to the five information types, producing five sets of unapproved documents. Each of these sets contains documents without the sentences related to a given information type. However, when explaining the core content of an information type, submitters may also write
additional sentences near the related sentences to introduce the
context or background.  We assume that for missing information types,
submitters would also omit such sentences.  Therefore, in all the documents, we removed both
the previous sentence and the next sentence of a sentence related to
an information type (i.e., the context of a sentence as explained in Section~\ref{sec:candidates}).
Additionally, we manually
checked the documents to remove sentences that indirectly indicate the
existence of a given information type. Such sentences are usually headings in the table of contents or pointers to specific sections that mention the name of the information type. For example, for \emph{issue
  condition and procedure} (T5), a prospectus may contain the sentence
``shareholders should consult the Chapter \emph{How to Subscribe For
  Shares}'', which determines the position of T5 in the text.  
  Lastly, we built a new test set containing 50 testing documents, ten of which are incomplete in terms of information types.

We ran \approach on the new test set using the same cross-validation setting as in RQ1.
When \approach reports that it did not find any sentence related to some information types, regulators are warned of missing information types.
We defined $\mathit{Precision}=\frac{|\mathit{TP}|}{|\mathit{TP}|+|\mathit{FP}|}$
and
$\mathit{Recall}=\frac{|\mathit{TP}|}{|\mathit{TP}|+|\mathit{FN}|}$,
where TP means \approach correctly identifies a missing information type (e.g.,  no related sentence is recommended);
FP represents the case in which \approach reports a missing information type but the document contains some related sentences;
FN corresponds to the case in which  \approach recommends sentences for a missing information type.
We also calculate \emph{F$_1$-score} according to precision and recall.
The definitions of evaluation metrics for compliance checking are different from those used in RQ1.
In this RQ, we assess the accuracy of \approach on finding information types instead of finding concrete sentences.
We focus on two cases, namely \approach cannot find any sentence or find at least one sentence related to an information type.

\subsubsection{Results}

\begin{table}[tb]
	\caption{Accuracy of \approach in Detecting Missing Information Types}
	\centering
\pgfplotstabletypeset[
	every head row/.style={
		before row={\toprule
},
		after row=\midrule,
	},
	every last row/.style={
		after row=\bottomrule
	},
	every nth row={5}{before row=\midrule},
	columns/ID/.style ={column name=ID}, 
	columns/P/.style ={column name=Precision}, 
	columns/R/.style ={column name=Recall},
	columns/F/.style ={column name=F$_1$-score},
	col sep=comma,
	string type,
]{data/rq-inspection.csv}

 	\label{tab:rq-inspection}
\end{table}

As shown in Table~\ref{tab:rq-inspection}, \approach achieves a precision ranging from \numrange{\MissPLow}{\MissPHigh} and a recall from \numrange{\MissRLow}{\MissRHigh} when detecting missing information types on the new test set. 
For T1 to T4, \approach identifies between 70\% and 90\% of the missing information types.
In contrast, the recall value of T5 is low. We speculate that this is caused by the
high number of sentences (on average, 36.4) in the training set that are related to T5:
many of the annotated sentences may discuss the general background of T5 instead of its core investment information. When using such general sentences to analyze a new prospectus, \approach could wrongly consider similar sentences in other locations as related; hence no warning of missing information types is reported.
Regarding precision, on average 84\% of the reported missing information types are correct.

\emph{The answer to RQ5 is that \approach detects 78\% of the missing
  information types with an average precision value of {\MissPAvg},
 which is a first significant step towards the semi-automated compliance checking of financial documents.}

\subsection{Threats to Validity}
\label{sec:threats}

One threat relates to the generality of the study. To address this
threat, we chose the content requirements  for UCITS prospectuses as
it is a representative investment regulatory framework that has
managed over \euro{10} trillion of assets across the world for the
past 30 years. We evaluated \approach with 100 prospectuses from
different investment companies.
These prospectuses are representative since they have been published
over many years and show a large diversity both in terms of number
of pages and number of sentences.
The difference in writing styles and document structures demonstrates the accuracy of \approach even in the context of widely varying financial documents.
Further, we selected a subset of five representative
information types due to the time required to annotate documents and
build the training set.  To increase generality, we selected these
information types by considering their importance, complexity, and
diversity.
These information
types are diverse in terms of the average number of related sentences.
They are also diverse in terms of the wording and writing styles as explained in Section~\ref{sec:eval:dataset}.
Therefore, \approach is expected to behave the same way with  other requirements
that have similar levels of details (i.e., from a single sentence to
several pages of sentences) as the five information types considered
in this work.
Using \approach for tracing
other information types requires domain experts to annotate the related
sentences for the information types in a set of  documents
that will constitute the training set of the ML model, and collect the key-phrase lists during the annotation.

Another threat relates to the process of creating the dataset.
The annotation of information types is a subjective process. Hence,
the annotation was conducted by three domain experts.
They annotated the ground truth independently and discussed possible inconsistencies to mitigate the subjectivity from a single domain expert.
Meanwhile, to answer RQ4, given the unavailability of unapproved prospectuses due to confidentiality reasons, we simulated unapproved prospectuses  by
removing related sentences and any sentence indicating the existence of information types.
The documents resulting from such a process may be different from actual, unapproved prospectuses, since the causes leading to a failed approval of a prospectus could manifest in several ways (e.g., with the omission of larger blocks of texts instead of only of the related sentences). Such potential differences could affect the accuracy of \approach when identifying missing information types. In the future, we plan to use actual unapproved prospectuses to evaluate \approach.

Third, the effectiveness of \approach may be affected by the
  presence of nuanced sentences in financial documents;
this implies the meaning of a sentence could change when only a specific word or the context is different.
To better analyze the meaning of different sentences, \approach mitigates this threat in two ways.
  First, as described in Section~\ref{sec:fine-grained}, \approach uses word importance similarity to identify the important words.
By analyzing the frequency, specificity, and universality of words,
\approach is expected to identify some unique words for a given information type.
Second, \approach integrates the domain knowledge of experts with two phrase lists (i.e., $\mathit{rlist}$ and $\mathit{ulist}$).
These lists contain the keywords used by domain experts to search the
possible location of related sentences; they are expected to help
\approach identify the  differences between sentences.

Finally, since regulations can change over time, it could be that the
language of the prospectuses in our dataset could change
significantly depending on the issue date of the documents, affecting the performance of \approach. We mitigated
this threat by assembling an experimental dataset that is not affected
by the changes in the law. More specifically,
all the information types selected by the domain experts for this
study have been required since the establishment of the UCIST Law in 2010, and
have not been affected by law changes since then.

\section{Discussion}
\label{sec:disc-pract-impl}

\subsubsection*{Inspection of financial documents.}
\label{sec:disc-inspection-fin-doc}
\approach can help regulator agents manually inspect
financial documents.

Given a new financial document, \approach can
analyze it and warn regulator agents of possibly missing
information types.
According to the results in RQ5, on average
\approach can find 78\% missing information types with a precision value of 84\%.
Regulator agents could then browse the document to
confirm the warnings.

Besides, \approach can also help regulator agents manually investigate financial documents.
Given a financial document, \approach can find 64.6\% sentences related to an information type with an average precision value of 82.4\% as shown in RQ1.
When \approach suggests these related sentences,
regulators can quickly read this small set of sentences (usually contains less than 50 sentences, depending on the average
number of related sentences in labeled documents) and check more carefully the sentences related to information types to refine their analysis. Thanks to its high
precision, in most cases \approach
can help regulator agents locate the right position of the sentences
related to a certain information type.

As part of future work, we plan to conduct a user study to analyze the
effect of \approach on reducing the inspection time of financial
documents.

\subsubsection*{Change of regulations.}
The change of the national and international laws can affect \approach,
since the content requirements of prospectuses might change
substantially over time due to changes introduced in the law:
the model trained on labeled prospectuses
may not correctly identify the information types written according to the new laws.
This problem can be solved by re-annotating the prospectuses with the information types
affected by the new laws and then by retraining the model.

We remark that changes in the law are infrequent~\cite{chalkidis2021regulatory}. For example, all the information types (i.e., T1--T5) we assess in this study are required to be present in every prospectus since the establishment of the ``UCITS Law 2010''. Hence, the document
re-annotation and model re-training steps are not expected to occur
frequently.

Moreover, since our evaluation has shown that \approach can identify
information types accurately with a relatively small number of labeled documents, the re-annotation step resulting from changes in the law may not be such an impractical undertaking.

\subsubsection*{Large language models.}
\label{sec:dis-llm}

In this work, we have not considered large language models
(LLM)~\cite{Brown:20,10.1145/3649506} for two reasons.

First, although there exist a number of LLM-based open-source and commercial
solutions (e.g., ChatPDF\footnote{\url{https://www.chatpdf.com}},
PDFChat\footnote{\url{https://www.pdfgpt.chat}},
PDF2GPT\footnote{\url{https://pdf2gpt.com}},
PDFGPT\footnote{\url{https://github.com/bhaskatripathi/pdfGPT}}) that
allow a user to upload a PDF and ask questions about its content,
all\footnote{At the time of writing this article, the PDFGPT GitHub
  page reported an upcoming release with support for LLMs that could be
  run locally, such as Falcon, Vicuna, Meta LLaMA.}
of them ultimately rely on 3rd-party services (e.g., OpenAI
API\footnote{\url{https://openai.com/product}}) for processing the
content of PDF files and retrieving answers to questions. Since
financial documents like prospectuses are to be treated as confidential documents at
the time a regulator performs compliance checking (i.e., before the
associated fund and the corresponding documents are made available to
the public), the use of LLM-based services could raise confidentiality
issues until such services could be fully run locally.

Second, LLMs are not specifically trained for tracing content
requirements of financial documents. Specializing LLMs for this task
and for the financial domain~\cite{10.1145/3604237.3626869} requires tremendous computation resources and would
represent a major research endeavor by itself. Recently, some LLMs
specialized for the financial domain (such as CFGPT~\cite{li2023cfgpt} and
BloombergGPT~\cite{wu2023bloomberggpt}) have been released. However,
these models have been
evaluated only in terms of sentiment analysis, named entity
recognition, and summarization tasks; assessing their performance on
other tasks such as  tracing content
requirements of financial documents and developing dedicated prompt
engineering best practices are open problems.

 \section{Related Work}
\label{sec:related_work}

Our approach is related to work done in the areas of requirement
traceability, mining financial data, information extraction from
  regulatory documents, and information structure identification.

\subsection{Requirements traceability}
\label{sec:rw-rt}

Requirements traceability can be considered as a kind of information
extraction~\cite{grishman2019twenty}, which extracts entities (e.g., named
entities~\cite{li2020survey}), relations (the relationship between two
entities~\cite{jiang2007systematic}), and events (e.g., knowledge about
incidents~\cite{huang2019improving}) from the text
with either rule-based~\cite{robaldo2011from} or statistical~\cite{li2020survey} techniques.
However, these tasks usually focus on analyzing small pieces of text (e.g., conversations, newsgroups, and weblogs)~\cite{zhang2017tacred, rajpurkar2018know, weischedel2011ontonotes}.
Driven by the recent deep learning advances (such as
BERT~\cite{devlin2019bert} and SpanBERT~\cite{joshi2020spanbert}), the
effectiveness of information extraction  for analyzing complex
documents has also significantly improved.
For example, \citet{chalkidis2021regulatory} performed document-level analysis with a BERT model to extract regulation documents of companies
that are affected by a certain law.
BERT is also widely used to build legal information retrieval systems~\cite{sansone2022legal}
However, applying these techniques to a different area (such as
the financial domain) always requires some sort of fine-tuning with domain-specific datasets.
Since fine-tuning is sometimes unstable on small datasets (with less
than 10k training samples)~\cite{zhang2020revisiting}, these
techniques cannot be applied in our context.  We indeed need to account for the limited
size of domain-specific datasets, bounded by the high cost of
annotations, which must be performed by domain experts and cannot be,
for example, crowd-sourced.
Moreover, the work by  \citet{chalkidis2021regulatory} aims to
retrieve a set of documents that are relevant to a specific document
(e.g., an EU directive) from a pool of documents (e.g., national
laws) and assumes to have a mapping (i.e., a transposition relation)
between EU directives and national laws to define relevance for the
retrieval task, whereas \approach works at the
sentence-level and does not require any mapping between regulations
and prospectuses.
\citet{castano2022knowledge} retrieve legal documents by building the legal ontology.
Their approach progressively enriches the terminological knowledge related to a concept and uses the enriched terms to retrieve documents.
However, the legal expert is required to review the new terms,
while \approach is an automated approach.

Requirements traceability has also been studied in the area of software engineering~\cite{cleland2014software, amaral2023ml},
where many requirement artifacts are written in NL~\cite{boutkova2011semi, merten2016software, wang2020detecting}.
Existing work infers trace links between high-level NL requirements (e.g., regulatory code) and low-level NL requirements (e.g., requirement specifications and privacy policies).
The traceability task is usually recast into an IR problem: taking high- or low-level requirements as queries to retrieve related or similar sentences from low-level requirements~\cite{hayes2003improving}.
IR techniques including latent semantic indexing, thesaurus, and relevance feedback have been investigated for this task~\cite{hayes2006advancing}.
To address the term mismatch between high- and low-level requirements,
the domain ontology~\cite{guo2017tackling}, word embedding~\cite{torre2020ai}, and indicator term mining~\cite{cleland2010machine, wang2020detecting} methods
have been explored for better sentence matching.
In addition, for a certain type of artifact (e.g., privacy policies), the NL text can be visualized~\cite{pullonen2019privacy} or standardized with domain-specific languages~\cite{caramujo2019rsl} to improve its traceability.

This study focuses on the traceability of content requirements, a type of non-functional requirements.
The concept of ``content requirement'' was initially proposed in the development of content-intensive interactive applications~\cite{bolchini2007branding,bolchini2004goal} (e.g., Web sites).
For example, in the case of a museum Web site, content requirements might be: ``present details for each painting'' or ``present museum collection history''~\cite{bolchini2004goal}.
This concept has also been recently investigated by
\citet{ceci2024defining}, who defined content
requirements found in the law as ``\emph{deontic rules} requiring that
some information is contained within an official document''; this is
the definition of content requirement that we have considered in this work.
Although \citet{ceci2024defining}
took financial regulations as a case study to define a model for
content requirements, they only discussed its potential application to support
compliance checking;
they did not provide an approach to automatically identify content requirements for compliance checking.
In this work, we take financial documents as a case study to automatically trace content requirements.

Regarding information type identification in content requirements,
several studies focus on the analysis of privacy policies.
In different countries,
privacy policies are subject to compliance with the law (e.g., the General
Data Protection Regulation (GDPR) in Europe).
For example, GDPR specifies that the privacy policies should indicate ``\emph{from
	which source personal data originates, and if applicable,
	whether it came from publicly accessible sources}'' (Art. 14.2(f)).
\citet{torre2020ai} proposed an AI-assisted approach to trace the sentences related to these information types in privacy policies.
\citet{amaral2023ml} improved this approach to enable automatic compliance checking between privacy policies and GDPR.
In the aforementioned works, a typical requirement is usually one or two sentences in length~\cite{hayes2006advancing};
in contrast, in this work we have focused on complex NL artifacts (i.e., financial documents),
which have thousands of sentences.
Further, similar sentences may have different meanings when the context differs; this is not the case for many artifacts (e.g., privacy policies).
To address these unique challenges, we proposed \approach to fully
consider the context, the content, and indicator words of each sentence
for better tracing financial content requirements.

\subsection{Mining financial data}
Collecting financial data (e.g., financial news, annual financial reports) plays a
critical role in improving the quality of financial services and
minimizing the risks of financial activities (e.g., portfolio
selection~\cite{zhang2020cost}, stock trading strategy
analysis~\cite{nuij2013automated}, stock price movements prediction~\cite{shi2018deepclue, feng2021hybrid}).
Existing studies report that financial data from Web media (e.g.,
financial news and discussion boards) has become increasingly salient for analyzing stock markets~\cite{li2017web}.
\citet{arslan2021comparison} and \citet{fan2020fusing} cluster and classify financial news to help analysts capture the core events in news.

Data mining has been applied not only to financial data from Web
media, but also to financial documents (e.g., annual financial
reports, 10-K). \citet{li2016table} extract financial tables from
annual financial reports, and automatically classify them into income
statements, balance sheets, and cash flow.  Mining financial tables
has also enabled activities like financial data
cross-checking~\cite{li2020cracking}, key performance indicators
tracing~\cite{brito2019hybrid}, and financial fraud
detection~\cite{craja2020deep, ravisankar2011detection}.

As to the analysis of NL sentences in financial
documents, \citet{kumar2016experiments} present a system
AEFDT to identify financial named entities (e.g., amortization
expense, swing factor). \citet{azzopardi2018controlled} propose a
controlled NL to write financial statements for
financial service compliance checking. In contrast to these works, in this
paper we focus on the task of tracing content requirements in
financial documents, which is important for financial enterprises and
regulators to ensure the completeness of financial documents and further
enable automated compliance techniques.

\subsection{Information Extraction from Regulatory Documents}
\label{sec:rw-if-reg}

In the domain of building construction,
several techniques have been proposed to extract information from
construction regulatory documents, which are typically large documents that specify construction regulations.
The requirements in these documents can be classified into quantitative requirements and existential requirements.
Quantitative requirements define the relationship between an attribute of a certain building
element/part and a specific quantity value.
For example, ``Habitable rooms shall have a net floor area of not less than 70 square feet''.
Existential requirements require the existence of certain building elements/parts. For example, ``The unit (efficiency dwelling unit) shall be provided with a separate bathroom''.
Existing approaches extract and organize these regulatory sentences
into a computer-processable rule representation (e.g., a structural
tuple $\langle$Subject, Attribute, Value$\rangle$) for compliance checking with actual building designs.
\citet{zhang2016semantic}~\cite{el2011automated} designed a set of pattern-matching-based rules to match each sentence in construction regulatory documents for extracting structural tuples.
\citet{zhang2021deep} used deep learning and transfer learning to
extract semantic and syntactic information elements from building regulations.

In building construction, information elements to be extracted are usually words or phrases (e.g., subject and value).
Existing studies~\cite{el2011automated} aim to extract elements of a
structural tuples from sentences, and organize them as tuples for analysis.
In contrast, \approach analyzes the content requirements in financial documents.
Each content requirement can be related to diverse numbers of sentences (e.g., from 3 to 36 sentences).

\subsection{Information Structure Identification}
\label{sec:rw:isi-s}
In scientific articles,
content requirement analysis can be considered similar to the task of
\emph{information structure (IS) identification},
which determines the topic or focus of a sentence in a given context~\cite{postolache2005data}.
A typical application of IS identification is to analyze the information types of sentences in scientific articles.
\citet{guo2011weakly} annotated sentences in abstracts of scientific articles
with seven categories (background, objective, method, result, conclusion, related work, and future work);
they found that it is much faster for readers to understand
IS-annotated articles than unannotated ones.
Since scientific articles are always required to including these categories of content,
it is important to identify IS automatically.
Most works use
feature-based machine learning, such as SVMs and logistic regression~\cite{de2021comparing} for this purpose.
Various linguistic features
and learning strategies have been explored, including sentence similarity, adjacency, part-of-speech, and topics~\cite{guo2015unsupervised, seaghdha2014unsupervised}.
Using the output of IS identification,
\citet{huang2017disa} developed a scientific writing advisor,
which helps refine scientific articles by suggesting similar sentences in the same IS category.

\approach is different from IS identification approaches for three reasons.
First, \approach identifies information types in financial documents.
Features for existing IS identification tasks (e.g., sentiment indicators) cannot be applied.
Second, in financial documents, only a small number of sentences is related to an information type,
which makes it more difficult to identify when compared with other types of
text (e.g., those found in sections of scientific articles).
Third, \approach aims to check the content requirement completeness of financial documents with respect to the identified information types;
as discussed above, existing IS identification works have different objectives.

\section{Conclusion}
\label{sec:conclusion}

In this paper, we proposed \approach, an approach to automatically
identify content requirements in financial documents.
Our approach combines IR and ML, to conduct analysis at multiple levels of granularity on financial documents in order to understand the context, semantics, and indicator terms of every sentence.
Furthermore, \approach uses a heuristic-based sentence selector, which considers the information from IR, ML, and domain-specific phrases to trace text spans related to the information types specified in the content requirements.
We evaluated \approach by assessing its effectiveness in identifying information types based on 100 financial documents from different investment companies.
Evaluation results show that \approach can accurately retrieve a large percentage of sentences related to  information types, with an average precision value of \FITIPAvg.
 \approach can thus effectively inform regulators about potentially
 missing information types and assist them in inspecting financial
 documents.

 As part of future work, we plan to improve the performance of
   \approach in the problematic cases identified in
   Section~\ref{sec:rq-accuracy} and to
 investigate the applicability of \approach on other types of financial documents with different information types.
We also plan to conduct a user study to assess the effectiveness of \approach to support the compliance checking of real-world unapproved prospectuses.

\backmatter

\section*{Acknowledgement}
	This research was funded in whole, or in part, by the Luxembourg National Research Fund (FNR), grant reference NCER22/IS/16570468/NCER-FT. For the purpose of open access, and in fulfillment of the obligations arising from the grant agreement, the authors have applied a Creative Commons Attribution 4.0 International (CC BY 4.0) license to any Author Accepted Manuscript version arising from this submission.
	\\
	Lionel Briand was partially supported by the Research Ireland grant 13/RC/2094-2, and the Canada Research Chair and Discovery Grant programs of the Natural Sciences and Engineering Research Council of Canada (NSERC).

\section*{Declarations}

\begin{itemize}
\item Conflict of interest: The authors declare that they have no known competing financial interests or personal relationships that could have appeared
to influence the work reported in this paper.
\item Availability of data and materials:
The non-annotated prospectuses, the raw output of \approach for
  each prospectus file in the test set, and
  the experiment results are available
at \url{https://figshare.com/articles/dataset/FITI_Experiment_Results/24481498}.
The machine learning models and the annotated prospectuses
cannot be distributed due to confidentiality and intellectual
  property agreements.
\end{itemize}

\bibliographystyle{sn-basic}

\end{document}